\documentclass{raa}
\usepackage{graphicx, times, amsmath, natbib}

\def\kms{\ {\rm km\,s^{-1}}}
\def\hmpc{\, h^{-1}{\rm Mpc}}

\begin{document}

  \title{Clustering of K-band selected local galaxies}

  \volnopage{Vol.0 (200x) No.0, 000--000}
  \setcounter{page}{1}

  \author{Bin Ma\inst{1} \and Kelai Meng\inst{1} \and Jun Pan\inst{1} \and
    Jiasheng Huang\inst{2} \and Longlong Feng\inst{1,3}}
  \institute{The Purple Mountain Observatory, 2 West Beijing Road,
    Nanjing 210008, China\\
  \and
  Harvard-Smithsonian Center for Astrophysics, 60 Garden Street, Cambridge,
  MA 02138, US\\
  \and
 Department of Physics, Texas A\&M University, College Station, TX 77843, US}

  \date{Received~~2009 month day; accepted~~2009~~month day}

\abstract{
    We present detailed clustering analysis of a large K-band selected
    local galaxy sample,
    which is constructed from the 2MASS and the SDSS and consists of $82,486$
    galaxies with $10 < K <13.5$ and $0.01<z<0.1$. The two-point correlation
    function of the magnitude-limited sample in real space at small scales
    is well described
    by a power law $\xi(r)=(r/6.44\pm0.23)^{-1.81\pm0.02}$.
    The pairwise velocity dispersion is derived
    from the anisotropic two-point correlation function and find
    the dispersion $\sigma_{12}=685\pm 17\kms$ if its scale invariance is assumed,
    which is larger than values measured in optical bands selected galaxy samples.
    We further investigate the dependence of the two-point correlation function
    and the $\sigma_{12}$ on the $g-r$ color and the $K$-band luminosity, obtain
    similar results to previous works in optical bands.
    Comparing a mock galaxy sample with our
    real data indicates that the semi-analytical model can not
    mimic the $\sigma_{12}$ in observation albeit it can approximate the
    two-point correlation function within measurement uncertainties.
    }

\authorrunning{B. Ma et al.}
\titlerunning{Clustering of K-band selected local galaxies}

\maketitle
\section{Introduction}
\label{sec:intro}
The track of galaxy formation and evolution keeps still
one of the pivotal but intricate problems in modern astrophysical research.
While there are large portion of prominent results mainly
derived in optical bands, it is worth of addressing that
observations of galaxies at distinct wave bands actually depict
galaxies' intrinsic properties in different aspects, which is essential
to our attempt to an unbiased understanding of the formation and evolution
of galaxies. The Near-Infrared (NIR) observation
is of particular interests.
A K-band $(2.15\mu m)$ selected galaxy sample has several advantages in studying galaxy.
The K-correction in the K-band
is a relatively small and better understood quantity comparing with the optical K-correction,
galaxy's K-band light is 5 -- 10 times less sensitive to dust and stellar populations, and
moreover is independent of galaxy spectral types at $z < 1$
\citep[e.g.][]{MobasherEtal1986, CowieEtal1994}.
Therefore K-band observation enables a much homogeneous sampling
of galaxy types.
It is even advantageous that the K-band luminosity is tightly
correlated with galaxy's stellar mass, so that the analysis at K-band is a
better probe of galaxy properties and relevant evolution history related
to the stellar component than those measurements in B- and $r$-bands.

Studying clustering strength using the two-point correlation function (2PCF), of
galaxies at different redshift bins is one of the most
straightforward and effective methods of exploring how galaxy properties
are related to the underlying dark matter distribution and henceforth to
the macro history of galaxy formation and evolution. For example, the
clustering of galaxies selected at $3.6\mu m$ with median redshift
$z\sim 0.5$ offered by the
{\em Spitzer} Wide-area Infrared Extragalactic (SWIRE) survey
is directly compared to the results of local galaxy samples selected
at K-band centered at $2.15\mu m$ \citep{WaddingtonEtal2007}.
Unfortunately in contrast with prolific photometric galaxy surveys
available in NIR bands, spectroscopic information of these galaxies
falls behind,
it is not strange to see that majority of the clustering analysis of NIR
galaxies resorts to angular correlation functions $\omega(\theta)$
\citep[e.g.][]{BaughEtal1996, McCrackenEtal2000, RocheEtal2003,
WaddingtonEtal2007},
the spatial correlation function $\xi(r)$ is applicable only
occasionally \citep{CarlbergEtal1997}.

In principle one can invert the angular 2PCF to
the real space spatial correlation function through the
Limber's equation \citep[see p.~194 of][]{Peebles1980}. The problem is
the precision of the inversion relies significantly on
the radial selection function of galaxies and the assumptions
of small angle approximation and a power law $\xi(r)$, which introduces
apparent uncertainties to its interpretation \citep{BernardeauEtal2002},
e.g. setting up the empirical evolution model of the real space
two-point correlation function
by $\xi(r, z)=(r/r_0)^{-\gamma}\times (1+z)^{\gamma-3-\epsilon}$
\citep{PhillippsEtal1978}.

Thanks to the Two-Micron All Sky Survey
Extended Source Catalogue \citep[2MASS XSC][]{JarrettEtal2000} and the
Sloan Digital Sky Survey \citep[SDSS][]{YorkEtal2000}, we are able to construct
a large K-band selected local galaxy sample to measure spatial two point
correlation function, and thus to built clustering
reference of local NIR galaxies. By combining
two-point correlation functions in both redshift and real spaces, we
are also able to derive quantities on the galaxies' peculiar velocities
which are largely ignored in characterizing clustering
of galaxies.
Actually to appropriately descibe galaxy
distribution, we should
work in the phase space supported by both of the position
and the peculiar velocity, especially in the nonlinear regime.
Thus measurement of the pairwise peculiar velocity dispersion is robust, providing
an important statistics to identify galaxy populations and test galaxy formation models
\citep[e.g.][]{ZhaoEtal2002, YoshikawaEtal2003, JingBorner2004b,
SlosarEtal2006, LiEtal2007}.

A brief summary of the sample used is presented in Section 2,
as well as the estimation method of 2PCF, including technical
issues such as random sample construction,
correction to the fibre collision.
In Section 3 we present measurements of the full, flux-limited
sample in both redshift- and real-spaces, in together with the
redshift-space distortion
parameter and the pairwise velocity dispersion.
In Section 4 the full sample is divided into different sub-samples according to their
color and luminosity to
probe the clustering as functions of these properties.

Throughout this paper, galaxy distances are obtained from redshifts
assuming a cosmology with $\Omega_{m}=0.3, \Omega_{\Lambda}=0.7$,
quoted in units of $\hmpc$. Absolute magnitudes quoted for galaxies
assume $h = 1$ to avoid the $-5\log_{10}h$ factor.

\section{Samples and estimation procedure}
\subsection{the K-band sample of local galaxies}
Our K-band galaxy sample is selected from the 2MASS XSC. Redshifts
for this smaple are obtained from the SDSS. The two catalogues
actually have been combined by \citet{BlantonEtal2005} into the New
York University Value-Added Galaxy Catalog (NYU-VAGC) \footnote{{\tt
http://sdss.physics.nyu.edu/vagc}} to form a local redshift
sample(mostly below $z \sim 0.3$) \footnote{ The version of the
NYU-VAGC used in this paper is the SDSS DR6
\citep{Adelman-McCarthyEtal2008}.}, with a coverage of  9938 $deg^2$
for photometric imaging and 6750 $deg^2$ for spectroscopic
observation.

The completeness of the cross-matched 2MASS+SDSS Catalogue has been
extensively discussed \citep[e.g.][]{BellEtal2003}. Following these
practices, we select the galaxies by the extinction-corrected Kron
magnitudes in the range $10 \leq K \leq 13.5$, and redshifts in the
range $0.01 \leq z \leq 0.1$. In the apparent magnitude limits
adopted here, the bright end is chosen to avoid incompleteness due
to the large angular sizes of galaxies, while the faint end is such
to match the magnitude limit of 2MASS. We also use a low redshift
cut rejects galaxies with redshifts seriously affected by Hubble
flow. The final flux-limited sample, our main sample, accumulates
78,339 galaxies with redshifts in total, with a median redshift
$\sim 0.06$.

\begin{figure}
\begin{center}
\resizebox{0.6\hsize}{0.5\hsize}{\includegraphics{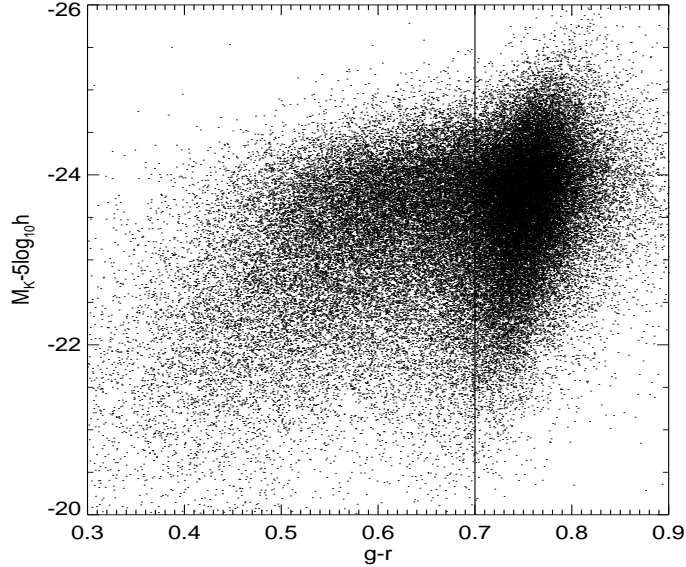}}
\end{center}
\caption{The K-corrected g - r color vs. K-band absolute magnitude for
all galaxies. The vertical line demarcates a simple cut at g - r =
0.7 }
\label{fig:cm}
\end{figure}

\begin{table}
\caption{The K-selected samples including both red and blue
populations. All samples are in the range of $10 < K < 13.5$ and
$0.01 < z < 0.1$. $M^*$ and $\alpha$ are fitted parameters of the
Schechter luminosity function \citep{Schechter1976}. $r_0$ and
$\gamma$ are obtained from fit to the projected 2PCF $w_p(\sigma)$.
The numbers of galaxies quoted here include those lack of redshift
due to fibre collision.}
\begin{center}
\begin{tabular}{ccccccc}
\hline
label & $N_{\mathrm{gal}}$ & $M^*$ & $\alpha$ & $r_0 (\hmpc)$ & $\gamma$ & $\beta$ \\
\hline
  Main & 82,486 & -23.38 & -0.94 & 6.44 $\pm$ 0.23 & 1.81 $\pm$ 0.02  & 0.39 $\pm$ 0.17 \\
  Red  & 46,689 & -23.31 & -0.48 & 7.62 $\pm$ 0.32 & 1.87 $\pm$ 0.03  & 0.35 $\pm$ 0.20 \\
  Blue & 35,797 & -23.08 & -1.11 & 4.92 $\pm$ 0.14 & 1.67 $\pm$ 0.02  & 0.50 $\pm$ 0.15 \\
\hline
\end{tabular}
\end{center}
\label{tab:fl}
\end{table}

The K-selected sample is further divided into
two sub-samples according to their SDSS color $(g-r)$: the blue sample with
$(g-r) < 0.7$ and red sample with
of $(g-r)>0.7$ (Figure~\ref{fig:cm} and Table~\ref{tab:fl}).
Both red and blue samples have similar space densities, but
galaxies in the red sample are systematically more luminous and therefore mainly inhabit
at higher redshift: the median redshift of red galaxies is $z_{med} \sim
0.065$ whereas for the blue $z_{med} \sim 0.054$.

A set of volume-limited sub-samples within different absolute magnitude
bins are also constructed for analysis (Table~\ref{tab:vl}).

\begin{table}
\caption{Volume-limited sub-samples and their 2PCF parameters.
${\bar n}$ is the mean number density in unit of $10^{-3} h^3 {\rm
Mpc}^{-3}$, and $r_0$ is in unit of $\hmpc$. All galaxy samples are
of $10 < K < 13.5$, $0.01 < z < 0.1$. $r_0$ and $\gamma$ are
obtained from $w_p(\sigma)$, the numbers in parentheses are values
of corresponding luminosity sub-samples of the mock galaxy
catalogue.}
\begin{center}
\begin{tabular}{c|ccccc}
\hline
Abs. Mag. & Redshift range & $N_{\mathrm{gal}}$ &
${\bar n}$& $r_0$ & $\gamma$
\\
\hline
 -25 -- -24 & 0.033 -- 0.100 & 14,152 & 0.87 & $7.17 \pm 0.30$ ($6.93\pm1.10$)
& $1.87 \pm 0.03$ ($1.82\pm0.07$)\\
 -24 -- -23 & 0.021 -- 0.067 & 19,714 & 3.83 & $5.52 \pm 0.28$ ($4.97\pm0.23$)
& $1.83 \pm 0.03$ ($1.87\pm0.03$)\\
 -23 -- -22 & 0.013 -- 0.042 & 8,969 & 7.08 & $5.23 \pm 0.65$ ($4.50\pm0.24$)
& $1.83 \pm 0.04$ ($1.86\pm0.04$)\\
 -22 -- -21 & 0.010 -- 0.026 & 2,232 & 7.19 & $4.46 \pm 0.57$ ($4.79\pm0.29$)
& $1.86 \pm 0.10$ ($1.92\pm0.04$)\\
\hline
\end{tabular}
\end{center}
\label{tab:vl}
\end{table}

\subsection{mock galaxy catalogue}
We also compare our sample with a mock galaxy catalogue to scout for
the performance of semi-analytical models for hierarchical galaxy
formation \citep[see][and references there in]{Baugh2006}. The mock
catalogue is derived from a high-resolution pure dark matter
cosmological simulation with $512^3$ particles in a box of size
$100\hmpc$ run by the GADGET2 code \citep{Springel2005}. The
fundamental cosmological parameters for the simulation are set as
$\Omega_\Lambda=0.762,\ \Omega_m=0.238,\ h=0.73, \ \sigma_8=0.74,\
n=1$, rest simulation parameters  are the same as  those  in the
pure dark matter run by \citet{LinEtal2006}, e.g. the particle mass
is $5.5\times 10^8 h^{-1} M_{\sun}$ and the soft length is $4.5
h^{-1}{\rm kpc}$. We generate the mock galaxies at $z=0.05$ from the
output of the simulation by the semi-analytical model of
\citet{KangEtal2005}.  Luminosities are the only derived property
for the mock galaxies, thus we divide the mock galaxies into a set
of luminosities bins and study galaxy and dark halo properties in
each bin.


\subsection{estimation of the two-point correlation functions}
\label{sec:method} Two types of 2PCFs for this sample are measured
to capture the clustering patterns: one is the isotropic $\xi(s)$
with $s$ denoting the separation of a pair of galaxies in redshift
space, and the other is the two-dimensional function $\xi(\sigma,
\pi=\sqrt{s^2-\sigma^2})$ with $\pi$ indicating the separation of a
pair of galaxies parallel to the line-of-sight and the $\sigma$
being the separation perpendicular to the line-of-sight. The latter
is mainly utilised to obtain the redshift distortion free function,
the projected 2PCF $w_p(\sigma)=\int \xi(\sigma, \pi){\rm d}\pi$,
and consequently the real space function $\xi(r)$ after proper
inversion \citep[e.g.][]{DavisPeebles1983, HawkinsEtal2003}.

The 2PCFs are measured using the
estimator of \citet{LandySzalay1993},
\begin{equation}
\xi=\frac{DD-2DR+RR}{RR}\ ,
\label{eq:LS}
\end{equation}
in which $DD$ is the normalised number of weighted
galaxy-galaxy pairs at given separation, $RR$ is the normalised number of
random-random pairs within the same separation in the random
catalogue and $DR$ is the normalised number of weighted galaxy-random
pairs. In general
the scale $s$ of $\xi(s)$ is binned uniformly in logarithm scale, and
for $\xi(\sigma, \pi)$, $\sigma$ and $\pi$ are binned in linear scale.

To proceed the estimation with Eq.~\ref{eq:LS}, an auxiliary sample
of completely random points in the exactly the same geometric window as
the galaxy sample is prerequisite.
The random samples should have the same redshift, magnitude and mask
constraints as the real data, with a smooth selection function
matching the $N(z)$ of the real data. The luminosity functions of the
flux-limited samples, used to generate the selection function, are
computed with the STY method \citep{STY1979} in form of the
Schechter function (Table~\ref{tab:fl}).
We generate
a random sample  ten times larger  than
the observed K-selected sample, and
the random samples  for red  and blue samples are 15 times larger than observed ones.

A weight is assigned to each galaxy and random point according to
their redshift and angular position to minimize the variance in
estimated $\xi$ \citep{Efstathiou1988, Hamilton1993},
\begin{equation}
\omega_i=\frac{1}{1+4\pi\,n(z)\Phi_i\,J_3(s)}\ ,
\end{equation}
where $\Phi_i$ is the selection function at the location of $i$th galaxy,
$n(z)$ is the mean number density, and $J_3(s)=\int_{0}^{s}\xi(s)s^2ds$.
The $J_3(s)$ is computed using a power-law $\xi(s)$ with correlation
length $s_0=12\hmpc$ and $\gamma_0=1$, we  find that the difference between our estimate and that
using a raw measured $\xi(s)$ is negligible, in agreement with
the conclusion of \citet{HawkinsEtal2003}. To normalize the pair
counts properly, we assure that the sum of weights of the
random catalogue equal the sum of weights of the real galaxy
catalogue, both are a function of scale.

We need to correct the  incompleteness in the spectroscopic sample
due to the effect of collided fiber constraints. The design of the
SDSS instrument means that fibers can not be placed closer than 55
arcsec on the same tile, members of a close pair of galaxies cannot
be targeted in a single fibre configuration so that the survey
misses a large fraction of close pairs. Because 2PCF will be
systematically underestimated without taking account of fiber
collisions effect, several methods are developed to correct this
bias \citep{ZehaviBlantonEtal2002, HawkinsEtal2003}. We adopt the
method of \citet{ZehaviBlantonEtal2002} by assigning the redshift of
the observed galaxy in a pair to the pair member whose redshift was
absent. Then we obtain ``collision corrected'' redshift. At large
scales, where both members of the pair contribute to the same
separation bin, this method is equivalent to double weighting. We
argue this method should perform  better on small scales because it
retains information about the known angular positions.
\citet{ZehaviEtal2005} showed that this method is an adequate
treatment: residual systematics for the redshift space correlation
function were considerably smaller than the statistical errors, and
this was even more true for the projected 2PCF $w_p(\sigma)$.

The covariance matrix is computed with the jack-knife technique
\citep{Lupton1993, ZehaviBlantonEtal2002}, the galaxy sample is splitted
into thirty separate regions of approximately equal sky area, and
then we perform the analysis thirty times, each time leaving a
different region out. The estimated statistical covariance of 2PCF
measured in two bins of $i$ and $j$ is then
\begin{equation}
{\rm Cov}(\xi_i,\xi_j)
=\frac{N-1}{N}\sum_{\ell=1}^{N}
(\xi_{i,\ell}-\overline{\xi_i})(\xi_{j,\ell}-\overline{\xi_j})\ ,
\end{equation}
in which $N$ is the number of jack-knife sub-samples.

\section{Clustering of the main sample}
\label{sec:results}
\subsection{2PCF in redshift space}
\begin{figure}
\resizebox{\hsize}{0.45\hsize}{\includegraphics{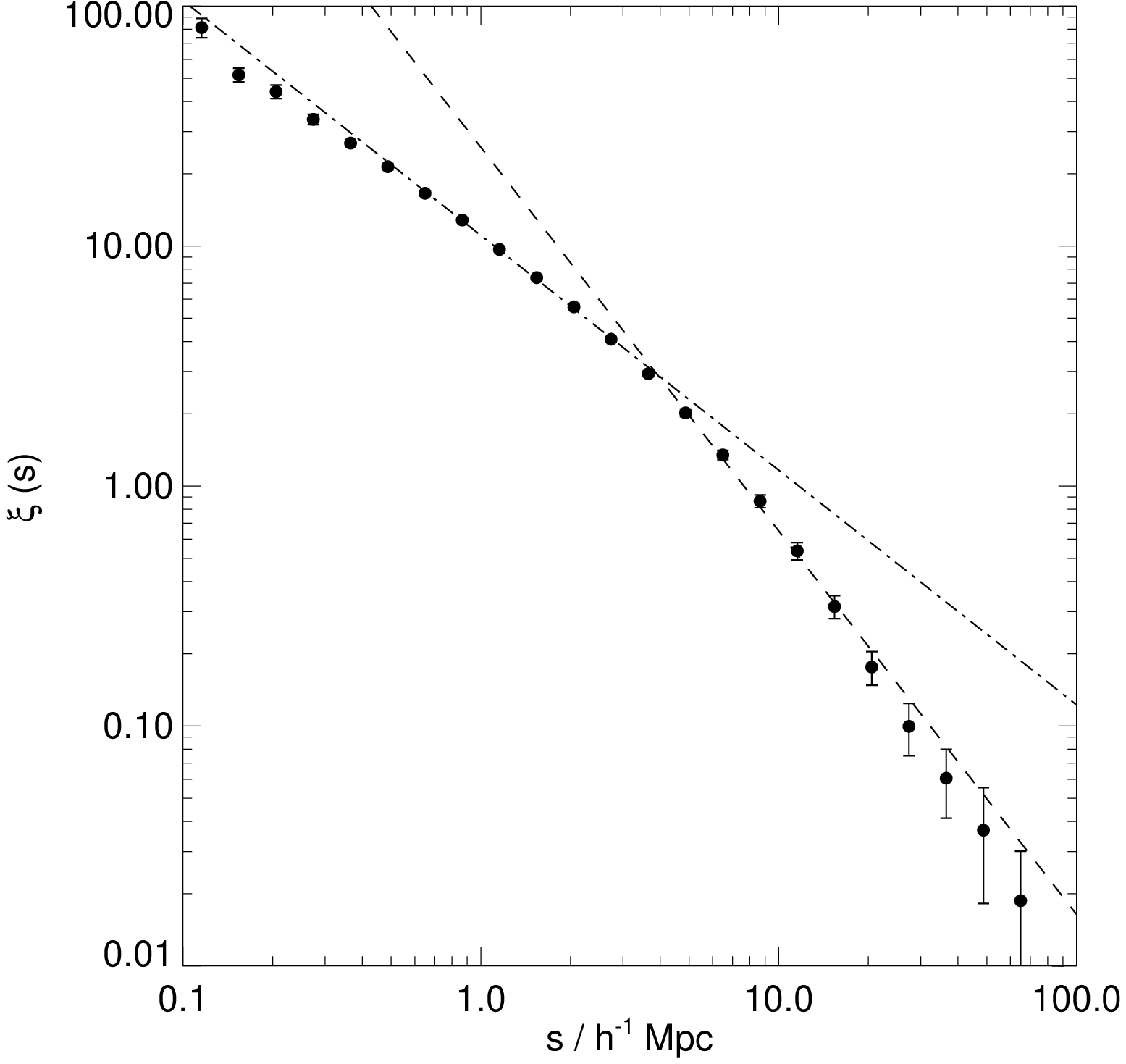}
\includegraphics{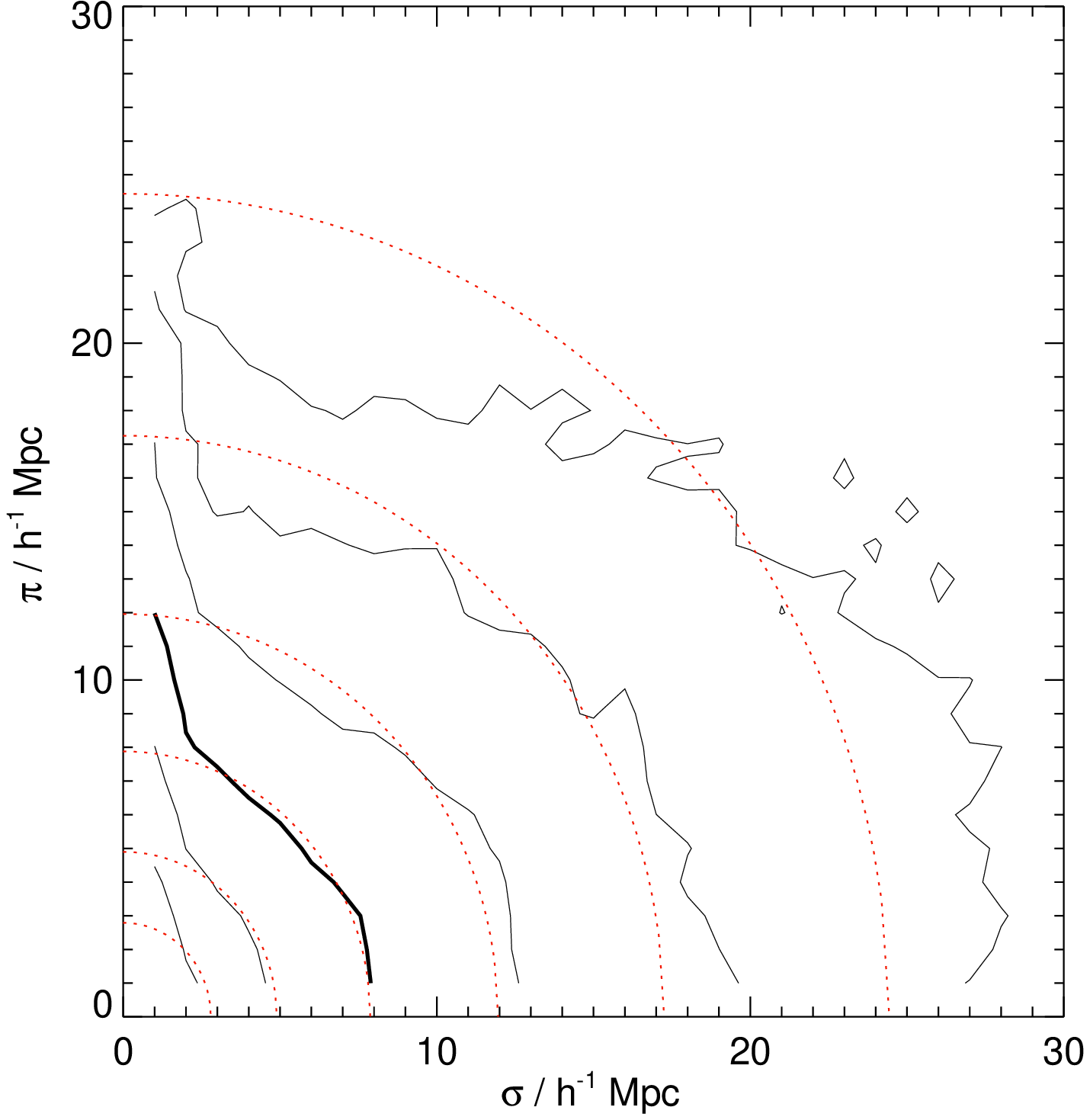}} \\
\resizebox{\hsize}{0.45\hsize}{\includegraphics{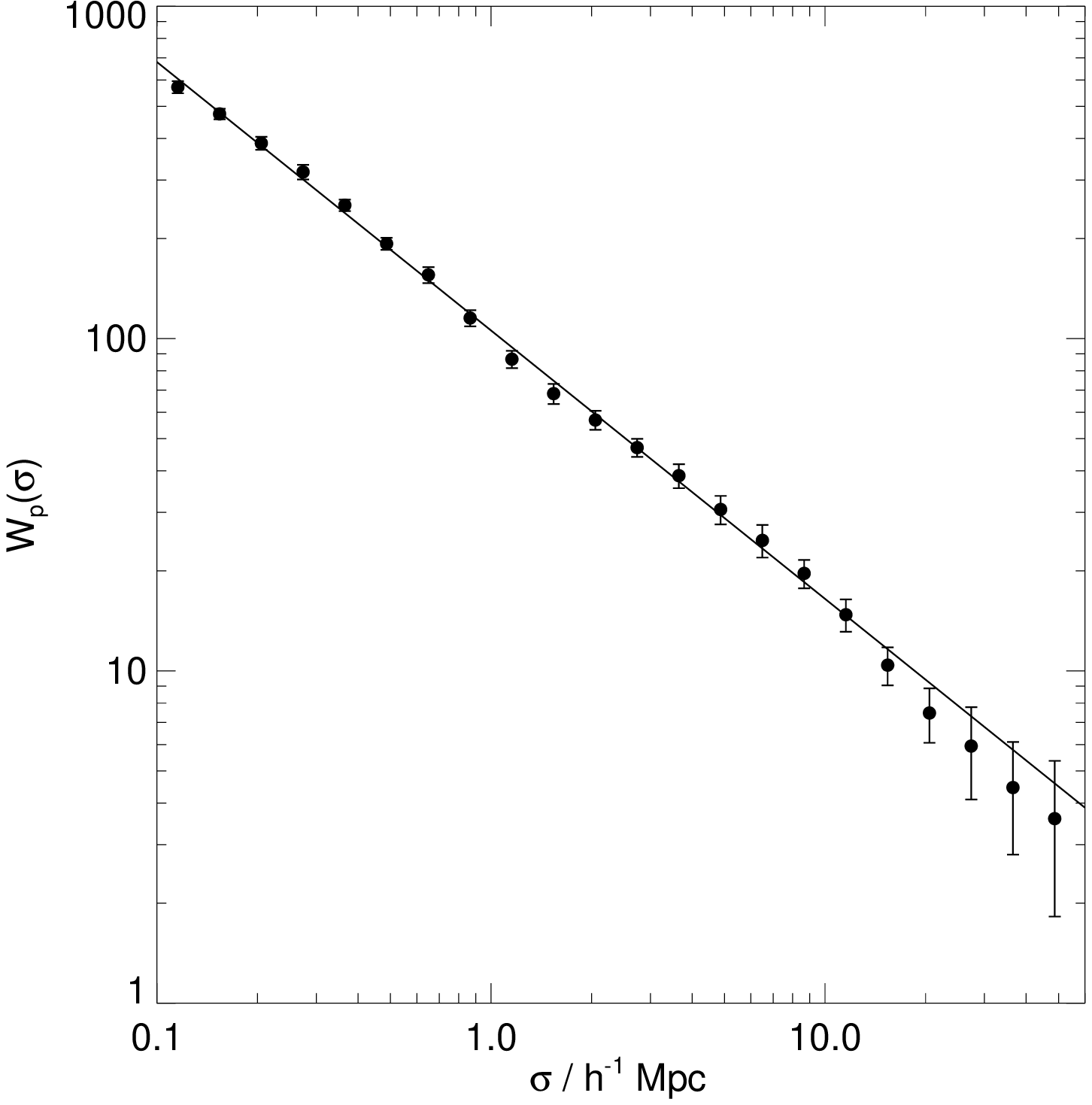}
\includegraphics{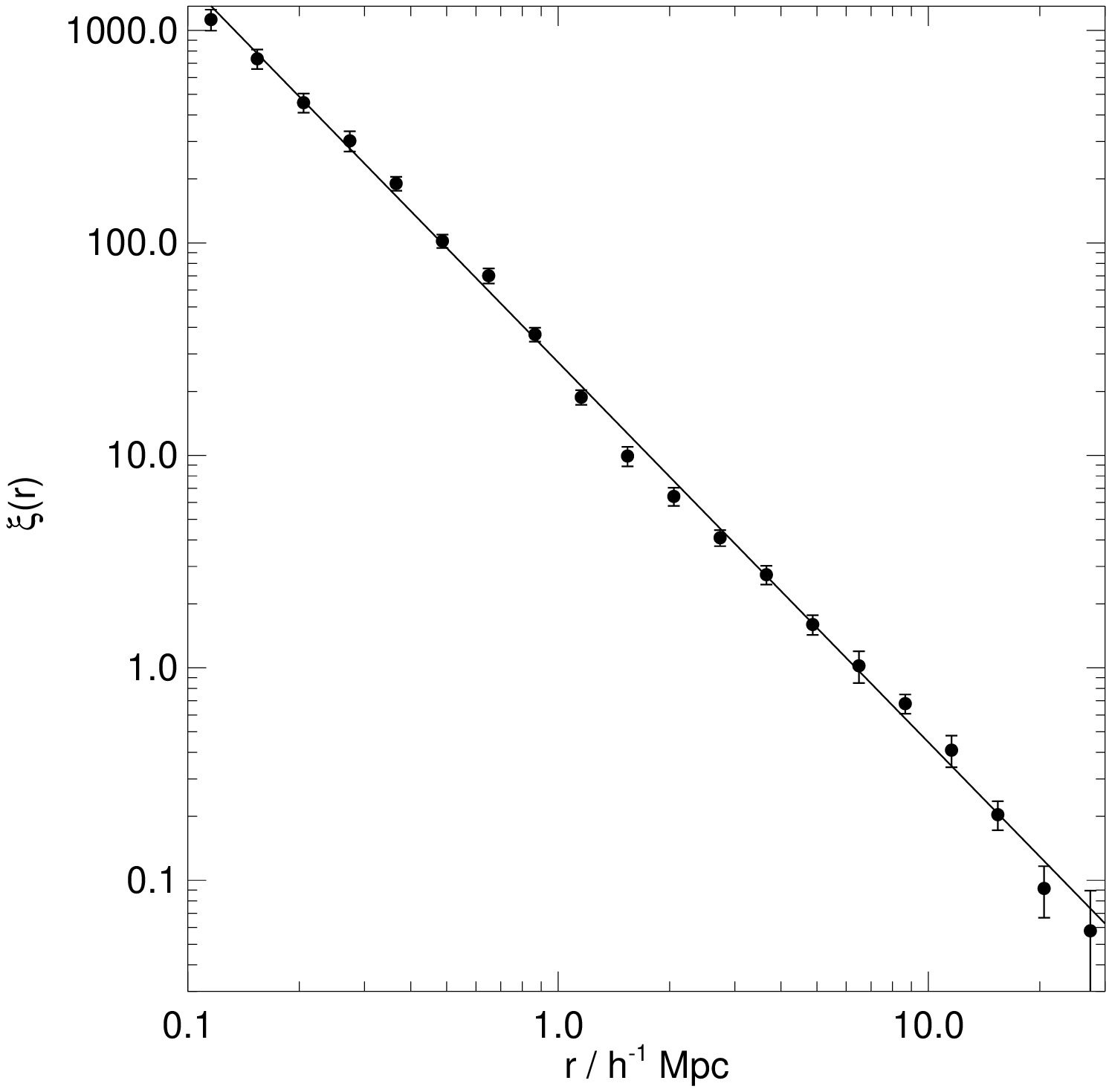}}
\caption{2PCFs of the main sample. Top-left: $\xi$ in redshift space,
the dot-dashed line is the power law at scales $0.1<s<3\hmpc$
and the dashed line is the fit at scales $3<s<10\hmpc$
(Eq.~\ref{eq:main_xis}). Top-right: $\xi(\sigma,\pi)$ contours,
$\sigma$ and $\pi$ are separations
perpendicular and parallel to the line of sight respectively, solid
contours are measurements at levels of $4, 2, 1, 0.5, 0.25, 0.125$ outwards
and dotted lines are the isotropic 2PCF expected in the absence of
redshift-space distortion.
Bottom-left: projected 2PCF $w_p(\sigma)$(solid
points), the solid line is the power law fit.
Bottom-right: 2PCF $\xi(r)$ in real space inverted from $w_p$ through
Eq.~\ref{eq:w2xi}, solid line is the power law transformed from
the fit to the $w_p$ on the left. Error bars in these plots are
the $1\sigma$ scatters of jackknife sub-samples.}
\label{fig:Wp}
\end{figure}

The first one we calculated is the 2PCF $\xi(s)$ for the main
sample from $s=0.01\hmpc$ to $s=100\hmpc$ in redshift space.
$s$ is divided into equally logarithmic bins of width
$\Delta\log_{10}(s/\hmpc)=0.125$.
$\xi(s)$ is not a single power law
at all scales (Figure~\ref{fig:Wp}), rather
\begin{equation}
\xi(s)=(\frac{s}{s_0})^{-\gamma_s},\ \ (s_0, \gamma_s)= \begin{aligned}
                  &(11.69\pm0.89,0.98\pm0.02), \ if\ s\in (0.1,3)\ ;\\
                  & (7.64\pm0.41,1.60\pm0.05), \ if\ s\in (3,10)\ .
                  \end{aligned}
\label{eq:main_xis}
\end{equation}

It is known that $\xi(s)$ also consists of the contribution from the galaxy
peculiar velocities causing the redshift distortion, in addition to the true spatial
correlation of galaxies \citep{Kaiser1987, Hamilton1998}. Thus
we need to break the degeneracy between
the spatial clustering and the velocity correlation before making direct
comparison of $\xi(s)$ in different redshift bins.

\subsection{The projected 2PCF}
Galaxy's peculiar motion only cause drifting of the radial position.
To minimize this effect, we can calculate the correlation function
as a function of $\sigma$ and $\pi$, where $\sigma$ is perpendicular
to the line-of-sight and $\pi$ is parallel to the line-of-sight.
Then the projection of $\xi(\sigma, \pi)$ onto the $\sigma$ plane is
independent to redshift space distortion and gives the information
of real space correlation function.

The effect of redshift distortion is clearly seen in the
$\xi(\sigma, \pi)$ contour of the main sample
in Figure~\ref{fig:Wp}: the contours
are elongated along the line-of-sight direction at small
$\sigma$ separation, exhibiting the phenomenon of fingers-of-God
by the random pairwise velocity; at large $\sigma$ scales,
the contours are squashed in the $\pi$ direction due
to gravitational coherent inflow
\citep{HawkinsEtal2003}.

Integrating the anisotropic $\xi(\sigma, \pi)$ over $\pi$
gives the projected 2PCF,
\begin{equation}
w_p(\sigma)=\int_{-\pi_{max}}^{+\pi_{max}}\xi(\sigma,\pi)d\pi
=\sum_i\xi(\sigma,\pi_i)\Delta{\pi_i}\ ,
\end{equation}
which has practically an upper limit $\pi_{max}=50\hmpc$. We tested
that there is little difference if using a larger cutoff.

$w_p$ is related to the 2PCF in real space through the Abel transform
\citep{DavisPeebles1983}
\begin{equation}\label{eq:xi2w}
w_p(\sigma)= 2\int_{\sigma}^{\infty}
\frac{r\xi(r)\,dr}{\sqrt{r^2-\sigma^2}}\ ,
\end{equation}
where the inversion renders $\xi(r)$  assuming a step function-like
$w_p(\sigma)=w_i$ at each bin centered at $\sigma_i$
\begin{equation}
\xi(\sigma_i)=\frac{1}{\pi}\sum_{j\geq i}
\frac{w_{j+1}-w_j}{\sigma_{j+1}-\sigma_j}\ln \left(
\frac{\sigma_{j+1}+\sqrt{\sigma^2_{j+1}-\sigma_i^2}}
{\sigma_j+\sqrt{\sigma_j^2-\sigma_i^2}} \right)
\label{eq:w2xi}
\end{equation}
for $r=\sigma_i$ \citep{SaundersEtal1992}. Mathematically the
inversion is not stable but in practice works well in $r<30\hmpc$.
If we simply assume $\xi(r)=(r/r_0)^{-\gamma}$ , the integral in
Eq.~\ref{eq:xi2w} can be done analytically, yielding
\begin{equation}
w_p=A\sigma^{1-\gamma},\ with \
A=r_0^\gamma\Gamma(\frac{1}{2})\Gamma(\frac{\gamma-1}{2})/\Gamma(\frac{\gamma}{2}),
\end{equation}
where $\Gamma$ is the Gamma function. It is true that
$w_p$ for our main sample
is a power law function at small scales (Figure~\ref{fig:Wp}),
the best-fit parameters are $\gamma=1.81\pm0.02$ and
$r_0=6.44\pm0.23$ in the regime of
$0.1\hmpc < \sigma < 18\hmpc$.

Table~\ref{table:comp} lists the results of our K-selected sample
and those selected in the other bands for comparison: $b_J$ of 2dFGRS and $r$ of SDSS. We
conclude that the correlation function will have larger
amplitude and steeper slope if the galaxies are selected at longer
wavelength band.

\begin{table}
\caption{Clustering of galaxies selected at different bands.
$r$-band (SDSS) result comes from
\citet{ZehaviBlantonEtal2002, ZehaviEtal2005} and the
$b_J$-band (2dFGRS) is given by \citet{HawkinsEtal2003}.}
\begin{center}
\begin{tabular}{ccccccc}
\hline
band & $N_{\mathrm{gal}}$ & $r_0$ ($\hmpc$) & $\gamma$ & $\sigma_{12}$ ($\kms$)\\
\hline
 Ks  & 82,486 &  6.44 $\pm$ 0.23 & 1.81 $\pm$ 0.02 & 685 $\pm$ 17\\
 r   & 154,014 &  5.59 $\pm$ 0.11 & 1.84 $\pm$ 0.01 & 640 $\pm$ 60 \\
 $b_J$ & 165,659 & 4.95 $\pm$ 0.27 & 1.72 $\pm$ 0.04 & 506 $\pm$ 52 \\
\hline
\end{tabular}
\end{center}
\label{table:comp}
\end{table}

\subsection{the pairwise velocity dispersion}
Currently there is no precise theory on
the full scale range redshift distortion
\citep{Scoccimarro2004}.
But we can still approach to this topic with reasonable assumptions.
Intuitively
redshift distortion can be approximated by certain convolution of
two components dominated in different regimes, coherent infall is
responsible for the clustering enhancement at large scales while
the smearing of correlation strength at small scales is attributed
to random motions.

\citet{Kaiser1987} found that at large scales the
boost to the power spectrum by the peculiar velocities takes a
particularly simple form, which was later completed and translated to
the real space by \citet{Hamilton1992},
\begin{equation}
\xi^\prime(\sigma,\pi)=\xi_0(s)P_0(\mu)+\xi_2(s)P_2(\mu)+\xi_4(s)P_4(\mu)\ ,
\label{eq:infall}
\end{equation}
where $P_\ell(\mu)$ is Legendre polynomials, $\mu=\cos(\theta)$ with
$\theta$ being the angle between $r$ and $\pi$. Assuming $\xi
=(r/r_0)^{-\gamma}$ then renders
\begin{equation}
\begin{aligned}
\xi_0(s)& =\xi(s)=\left(1+\frac{2\beta}{3}+\frac{\beta^2}{5}\right)\xi(r) \\
\xi_2(s)& =\left(\frac{4\beta}{3}+\frac{4\beta^2}{7}\right)
\left(\frac{\gamma}{\gamma-3}\right)\xi(r) \\
\xi_4(s)& =\frac{8\beta^2}{35}
\left(\frac{\gamma(2+\gamma)}{(3-\gamma)(5-\gamma)}\right)\xi(r)\ ,
\end{aligned}
\label{eq:kaiser}
\end{equation}
where $s$ and $r$ have the correspondent values, $\beta$ is the
linear redshift distortion parameter,
$\beta\approx\Omega_0^{0.6}/b$, and $b$ is the linear bias
parameter. The first equation is independent on the form of
$\xi(r)$.

To incorporate effects of random motion, the anisotropic 2PCF in
redshift space is then
approximated by the convolution of the $\xi^\prime(\sigma,\pi)$ by
Eq.~\ref{eq:infall} with the distribution function of the
pairwise velocity $f(v_{12})$ \citep[c.f.][]{Peebles1993},
\begin{equation}
\xi(\sigma,\pi)=\int_{-\infty}^{+\infty}
\xi^\prime(\sigma,\pi-\frac{v_{12}}{H_0})f(v_{12})dv_{12}\ ,
\label{eq:fog}
\end{equation}
and in general $f(v_{12})$ is assumed to be an exponential distribution of
dispersion $\sigma_{12}$
\begin{equation}
f(v_{12})=\frac{1}{\sigma_{12}\sqrt{2}}\exp
\left(-\frac{\sqrt{2} v_{12}}{\sigma_{12}}\right)\ .
\label{eq:fv}
\end{equation}

After measuring $\xi(s)$ and $\xi(r)$ at large scales, we
can obtain the redshift distortion parameter $\beta$ via the first
equation in Eq.~\ref{eq:kaiser} easily. Then we combine
Eq.~\ref{eq:infall} -- Eq.~\ref{eq:fv} to fit the
$\xi(\sigma,\pi)$ data grid to determine other model parameters.
However there are implicit assumptions in the
prescription: (1) the linear bias parameter $b$ is forced to be
scale-independent; (2) the pairwise velocity dispersion $\sigma_{12}$ is
presumed invariant to the separation $\pi$ along the line-of-sight but
could be a function of the perpendicular separation $\sigma$. Therefore
the resulting $\sigma_{12}$ in the model is not the actual true
pairwise velocity dispersion. To have meaningful comparison with
simulations, we need to estimate the $\sigma_{12}$ from the 2PCFs of
the simulation data in the same way as of the galaxy sample.

\begin{figure}
\resizebox{\hsize}{0.45\hsize}{\includegraphics{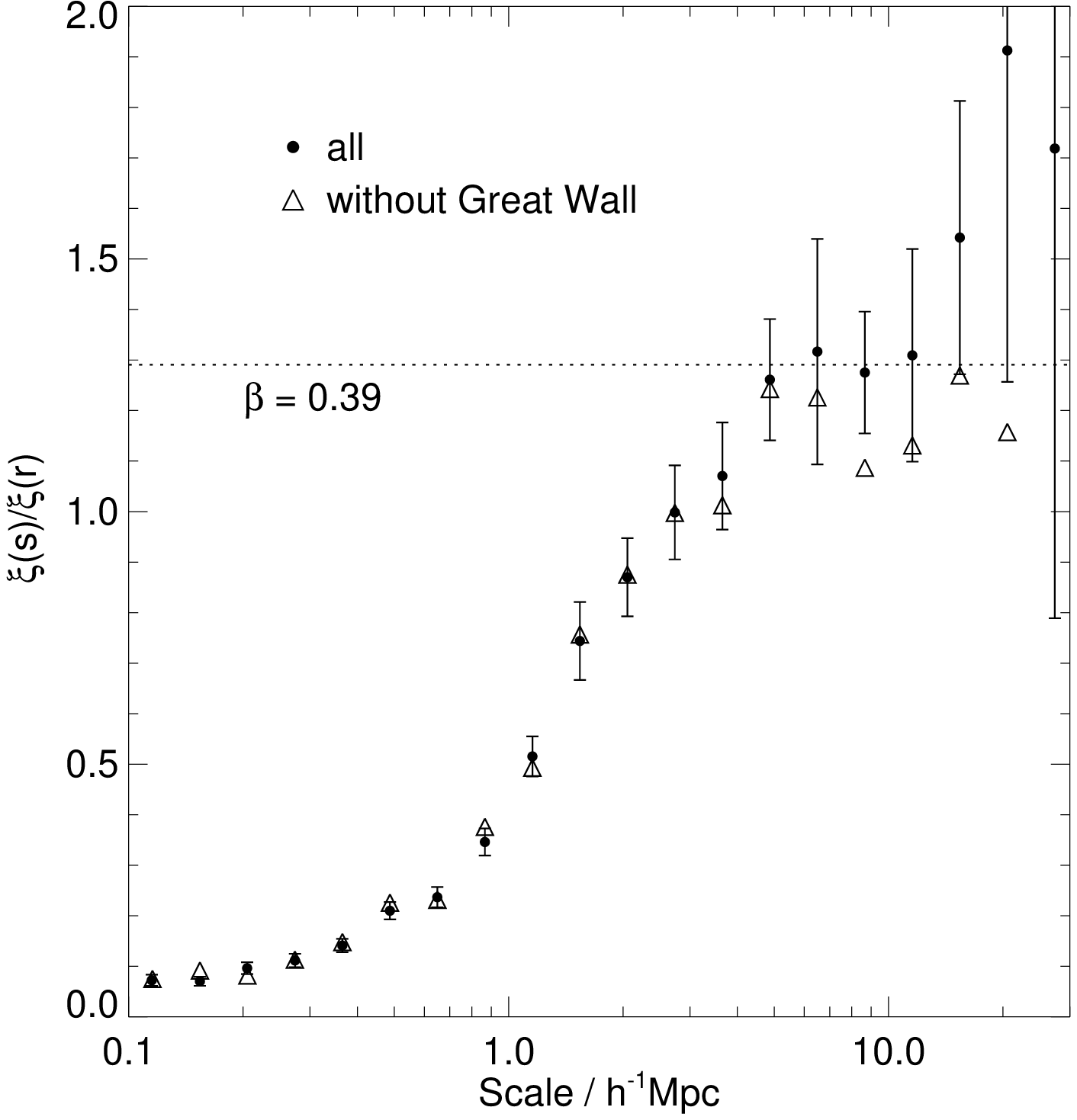}\includegraphics{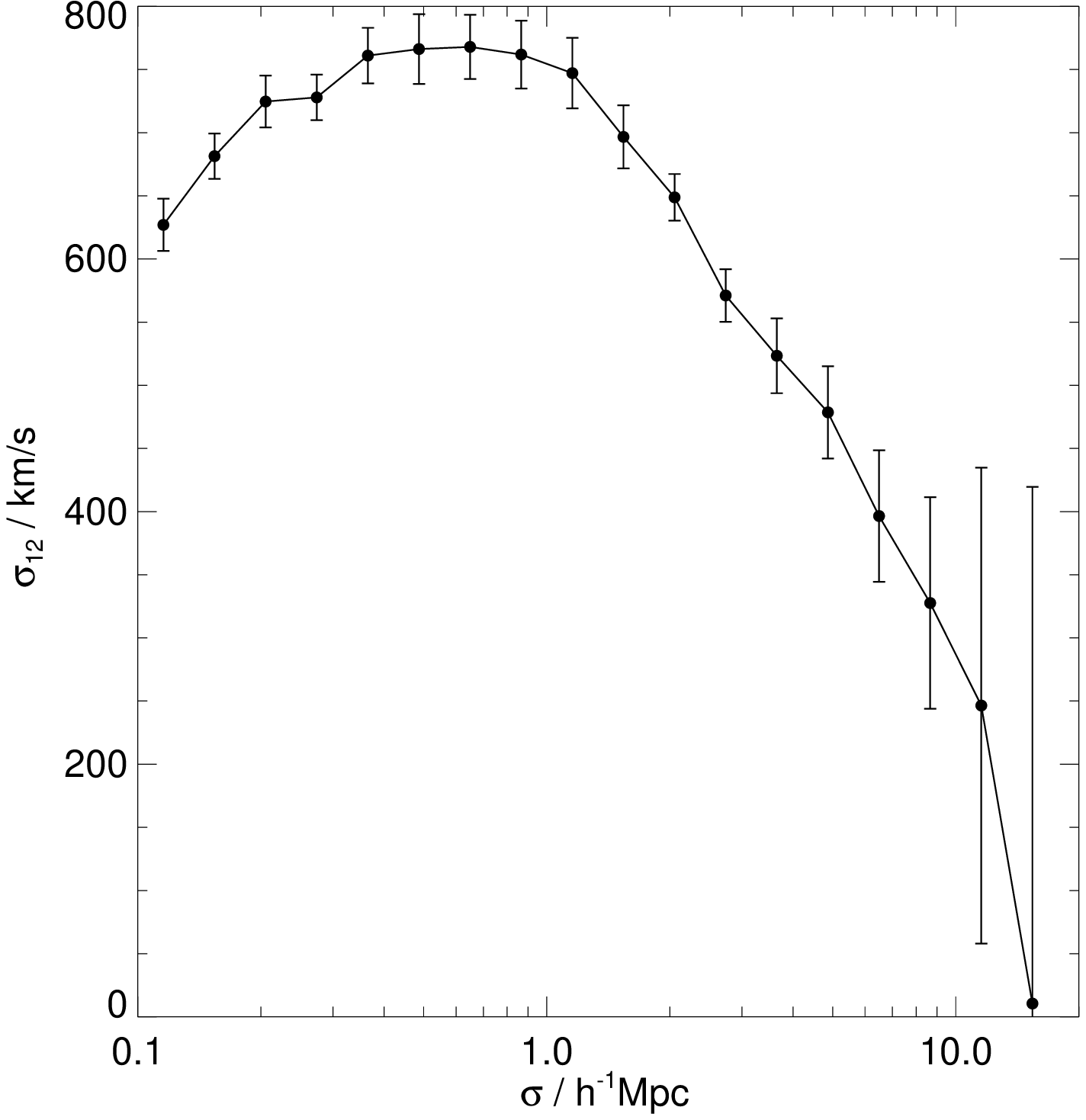}}
\caption{Left: ratios of $\xi(s)/\xi(r)$ of the full sample (solid
circles) in comparison with that of the sample without the Sloan Great Wall
(open triangle), the dotted line corresponds to $\beta =0.39$.
Right: the derived pairwise velocity dispersion $\sigma_{12}(\sigma)$.
Error bars are obtained by jackknife re-sampling.}
\label{fig:s2r}
\end{figure}

Figure~\ref{fig:s2r} shows the ratio of the redshift-space 2PCF to
the real-space 2PCF, and the derived $\sigma_{12}$ as function of
$\sigma$. $\xi(s)/\xi(r)$ increases with scales, and  becomes
roughly constant in the range $4\hmpc < r < 13\hmpc$ as expected,
however, the ratio increases again at larger scales. If we exclude
the galaxies in the Sloan Great Wall region ($11^h < {\rm R.A.} <
14^h$), the $\xi(s)/\xi(r)$ ratio indeed does not show the
up-shooting anymore. Nevertheless, there are no appealing arguments
about whether chopping off galaxies in the Sloan Great Wall
guarantees fairness, we just pick up points in the scale range $(4,
13\hmpc)$ to fit to a constant and get $\xi(s)/\xi(r)=1.29 \pm
0.14$, or $\beta = 0.39 \pm 0.17$, considering the fact that the
final $\sigma_{12}$ is not sensitive to the $\beta$ at all
\citep{LiEtal2006b}.

The $\sigma_{12}$ as a function of projected separation
$\sigma$ shown in Fig.~\ref{fig:s2r} has a classical
shape as many other measurements
\citep[e.g.][]{JingBorner2004b, LiEtal2007}:
$\sigma_{12}$ rises from $650\kms$ up to $750\kms$ as
$\sigma$ increases from $0.1\hmpc$ to $0.3\hmpc$, forms a plateau
till to $\sigma \sim 1\hmpc$, then drops down again at larger scales.
If a scale independent $\sigma_{12}$ is assumed, the best fitted
$\sigma_{12} = 685\pm17\kms$ within $0.1\hmpc < r < 18\hmpc$.
Our $\sigma_{12}$ are significantly higher than that of $b_J$-band sample
\citep[2dFGRS,][]{HawkinsEtal2003} and slightly larger than the
$r$-band sample \citep[SDSS,][]{ZehaviBlantonEtal2002}.

\section{color sub-samples}

\begin{figure}
\resizebox{\hsize}{0.45\hsize}{\includegraphics{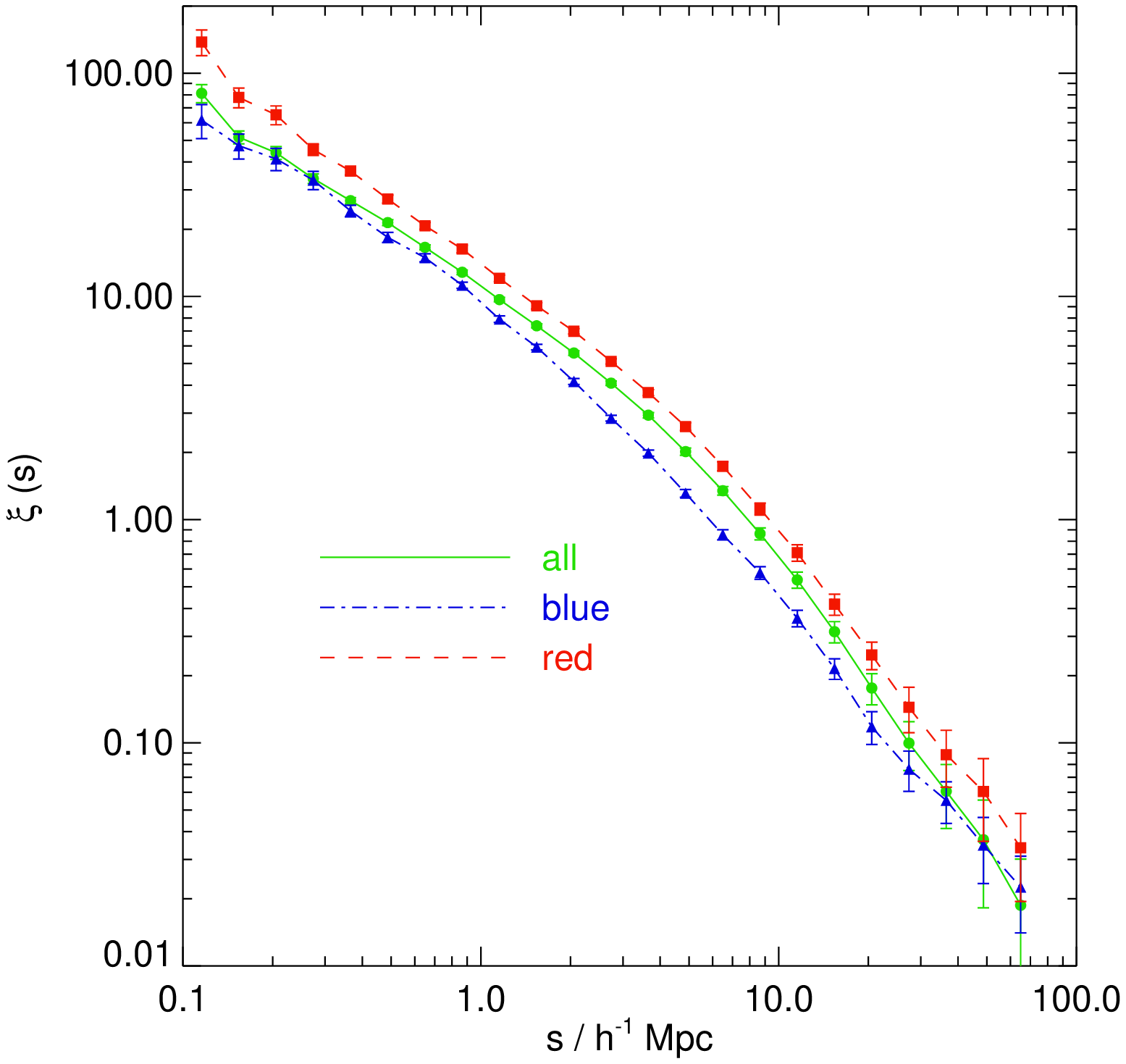}
\includegraphics{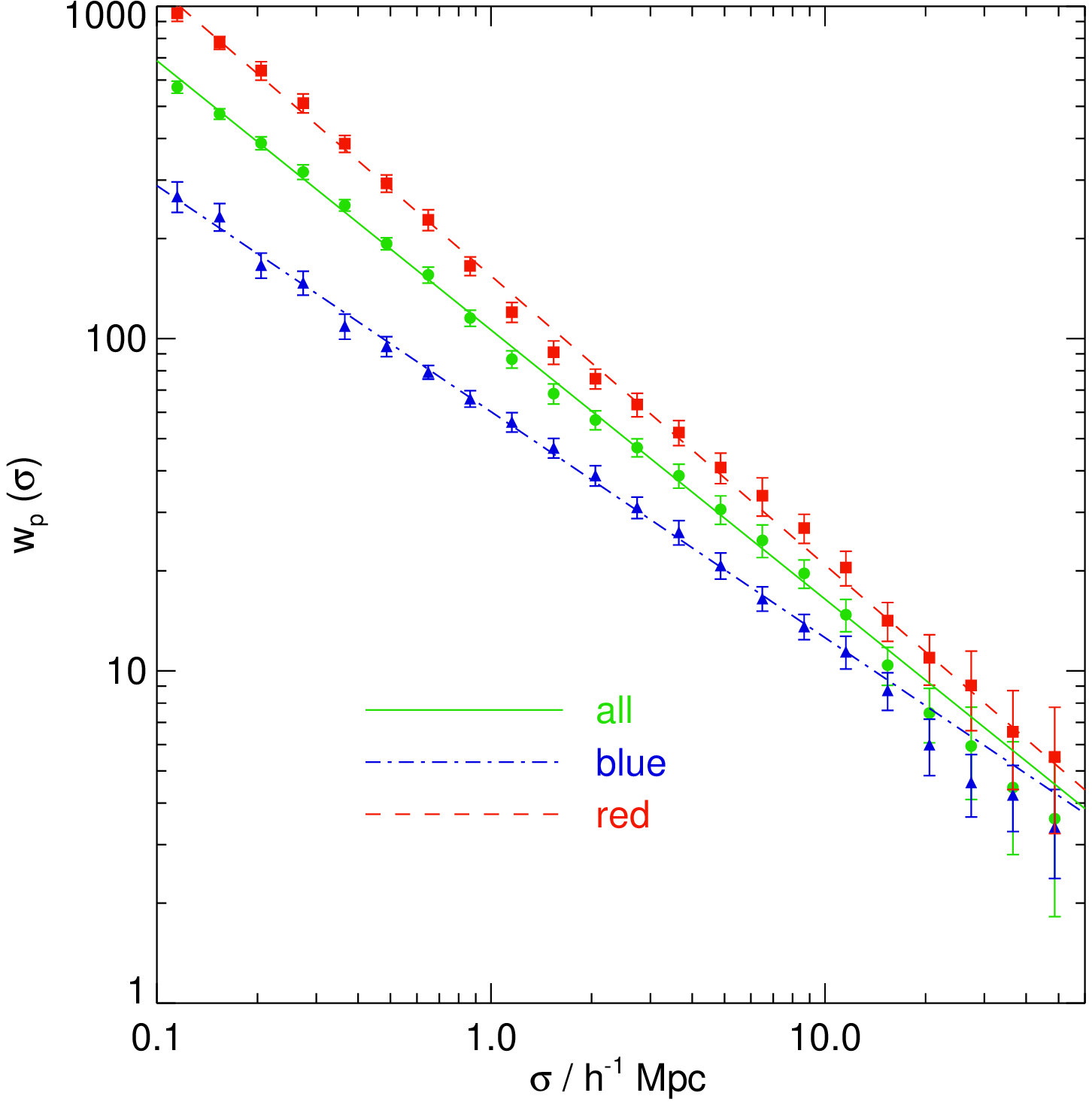}}\\
\resizebox{\hsize}{0.45\hsize}{\includegraphics{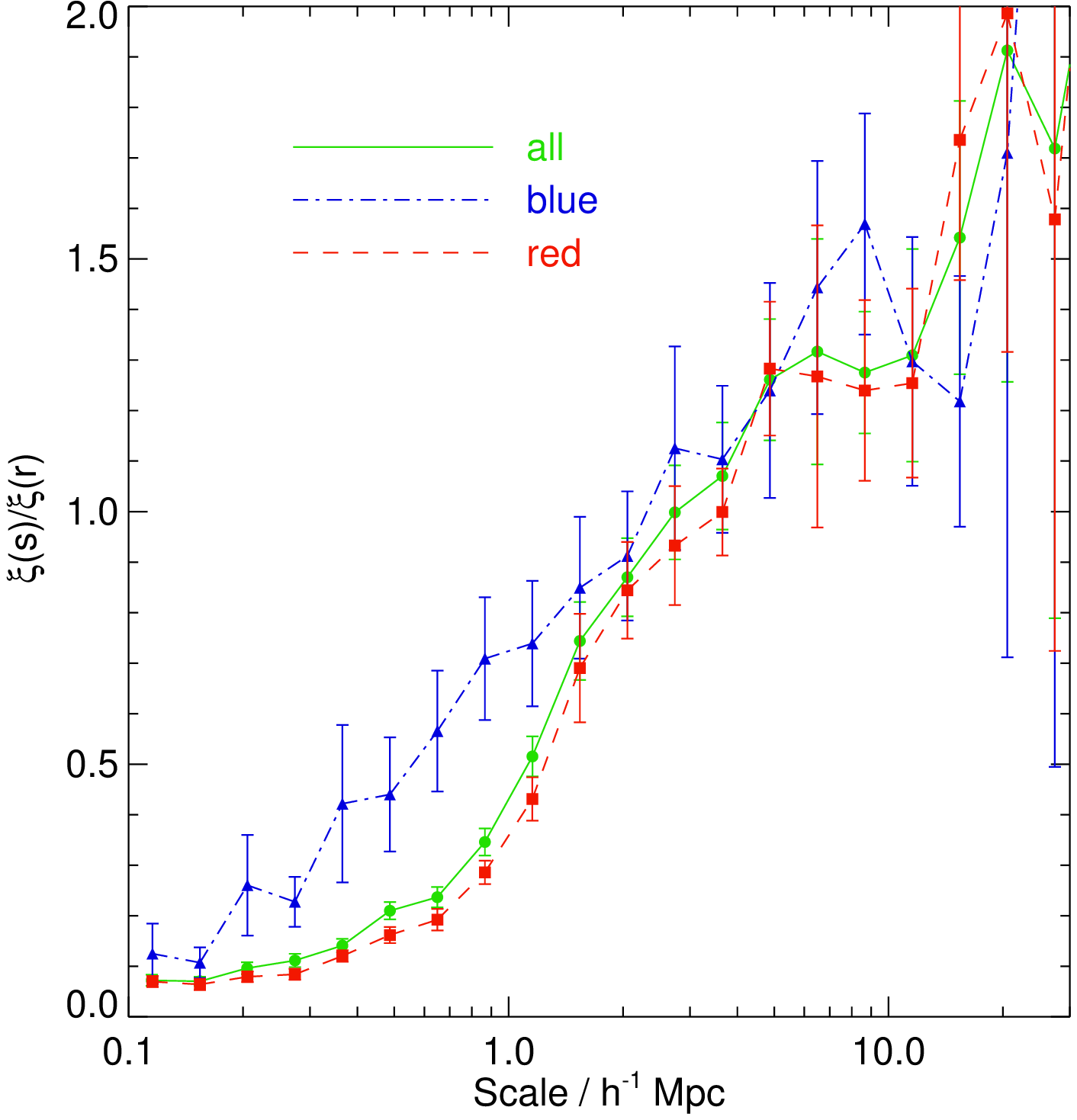}
\includegraphics{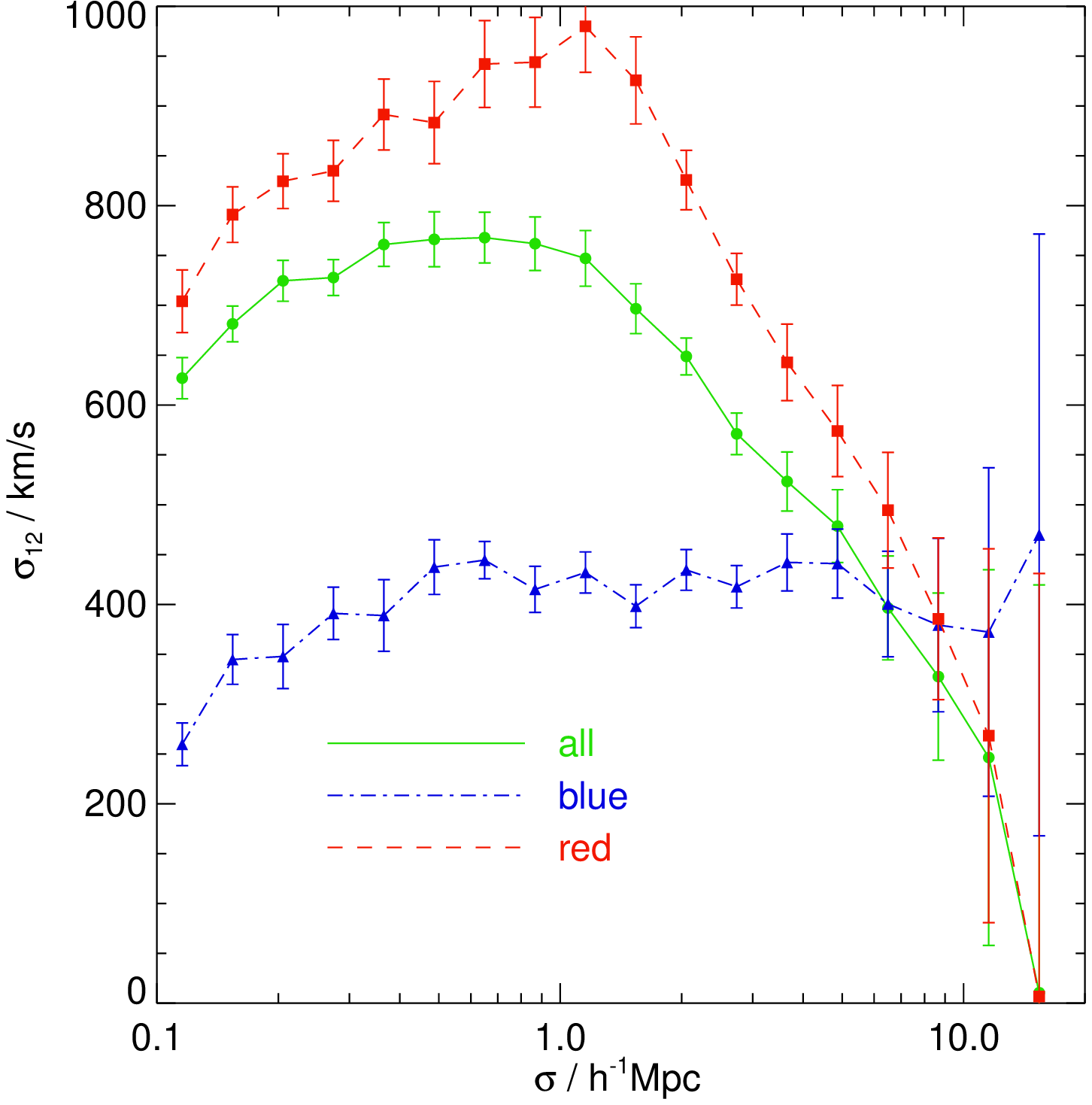}}
\caption{$g-r$ color dependence of 2PCFs and $\sigma_{12}$.
All error bars are derived from jackknife sub-samples.}
\label{fig:xis_color}
\end{figure}

Figure~\ref{fig:xis_color} displays the clustering dependence on the
color of galaxies, from $\xi(s)$ and $w_p$ as numerous works already
revealed  that spatial clustering for red galaxies is much stronger
than for blue galaxies (see also Table~\ref{tab:fl}). The slope of
the 2PCF is a direct indicator of strength of the galaxy dynamic
non-linearity, thus red galaxies with steeper 2PCF are more harassed
by the gravitational turbulence at small scales or local structures
than blue galaxies. In the scheme of halo models, the slope of 2PCF
is determined by the percentage of contribution from the one-halo
term and two-halo terms, rather than the mass of the halo in which
the galaxy resides \citep{CooraySheth2002}. A steeper 2PCF at small
scales contains more power from the one-halo term. If  galaxies are
simply centrals and satellites in dark halo, it leads to a
conclusion that the satellite fraction in the red sub-sample is
higher than in the blue sub-sample, although red galaxies incline to
occupy more massive halos than blue galaxies. In fact the
interpretation is confirmed by the velocity information:
$\sigma_{12}(\sigma)$ of blue galaxies is very flat and has much
lower amplitude than red galaxies over wide range of projected
separation, which is exactly what is observed in simulations when
reducing satellite fraction in mock galaxy samples
\citep{SlosarEtal2006}. We also note that the color dependence of
$\sigma_{12}$ is very similar to the simulation results of the old
and young populations galaxies demonstrated by
\citet{WeinbergEtal2004}.

\section{volume-limited sub-samples and comparison with mock catalogue}
We construct four volume-limited sub-samples to investigate the
luminosity dependence of clustering (Table~\ref{tab:vl}). The main
sample is divided to 4 absolute magnitude bins centered
approximately from $M^* + 2$ to $M^* - 1$, where $M^* =-23.38$ is
the characteristic luminosity of the Schechter function
\citep{Schechter1976}. The number density for galaxies in the lowest
luminosity bin is 8 times higher than that in the highest luminosity
bin.

\begin{figure}
\resizebox{0.5\hsize}{!}{\includegraphics{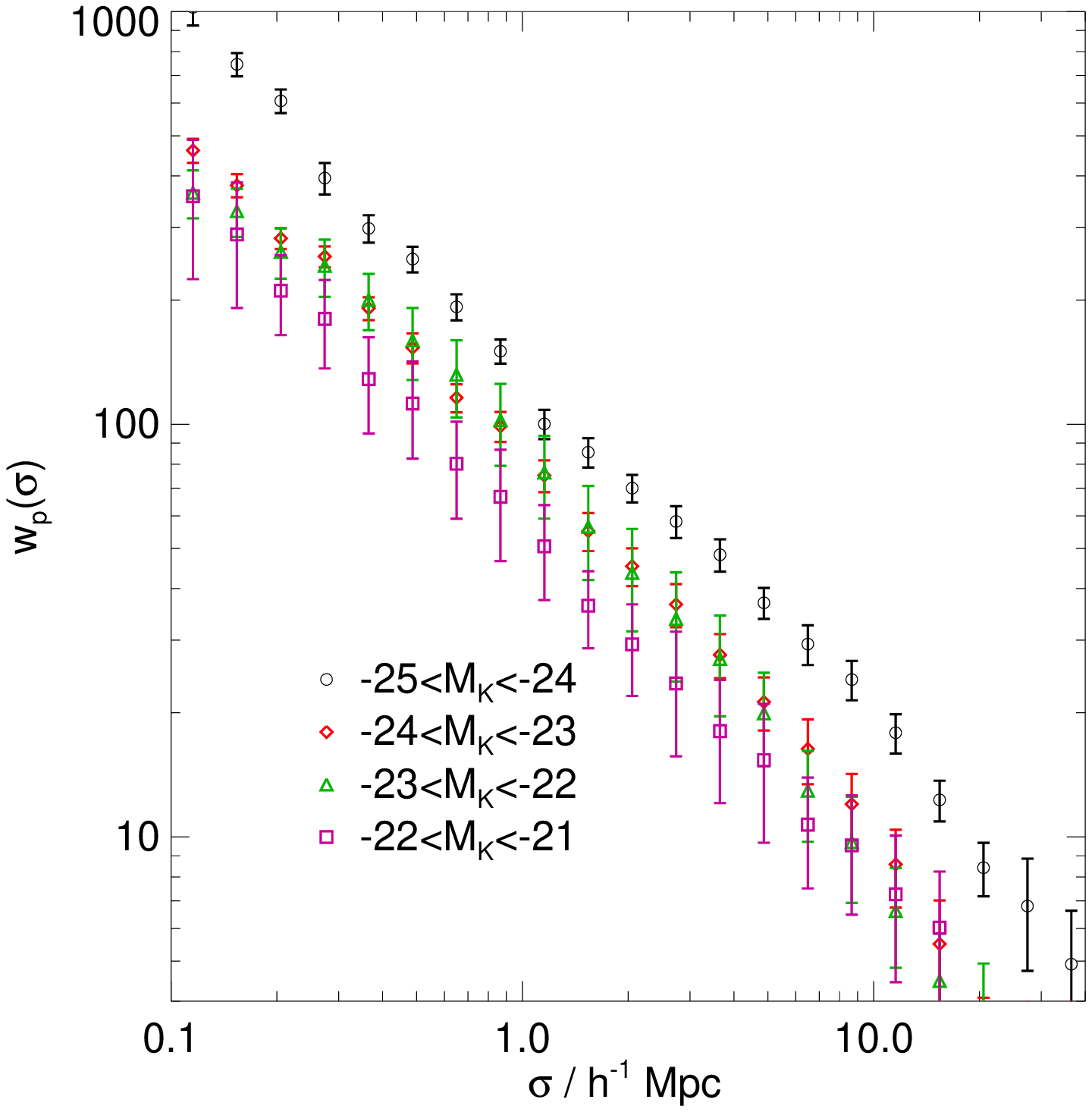}}
\\
\resizebox{\hsize}{!}{\includegraphics{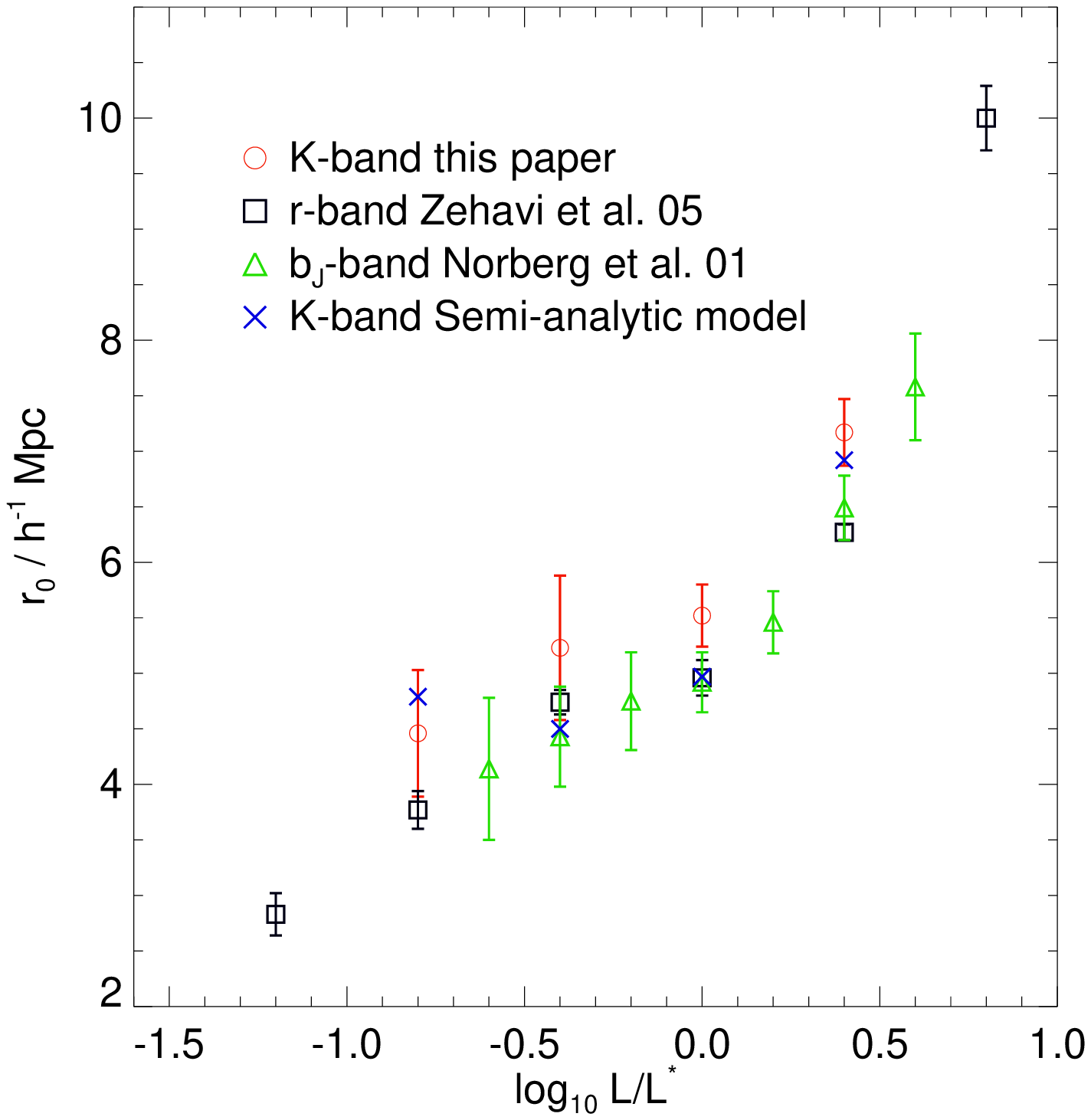}\includegraphics{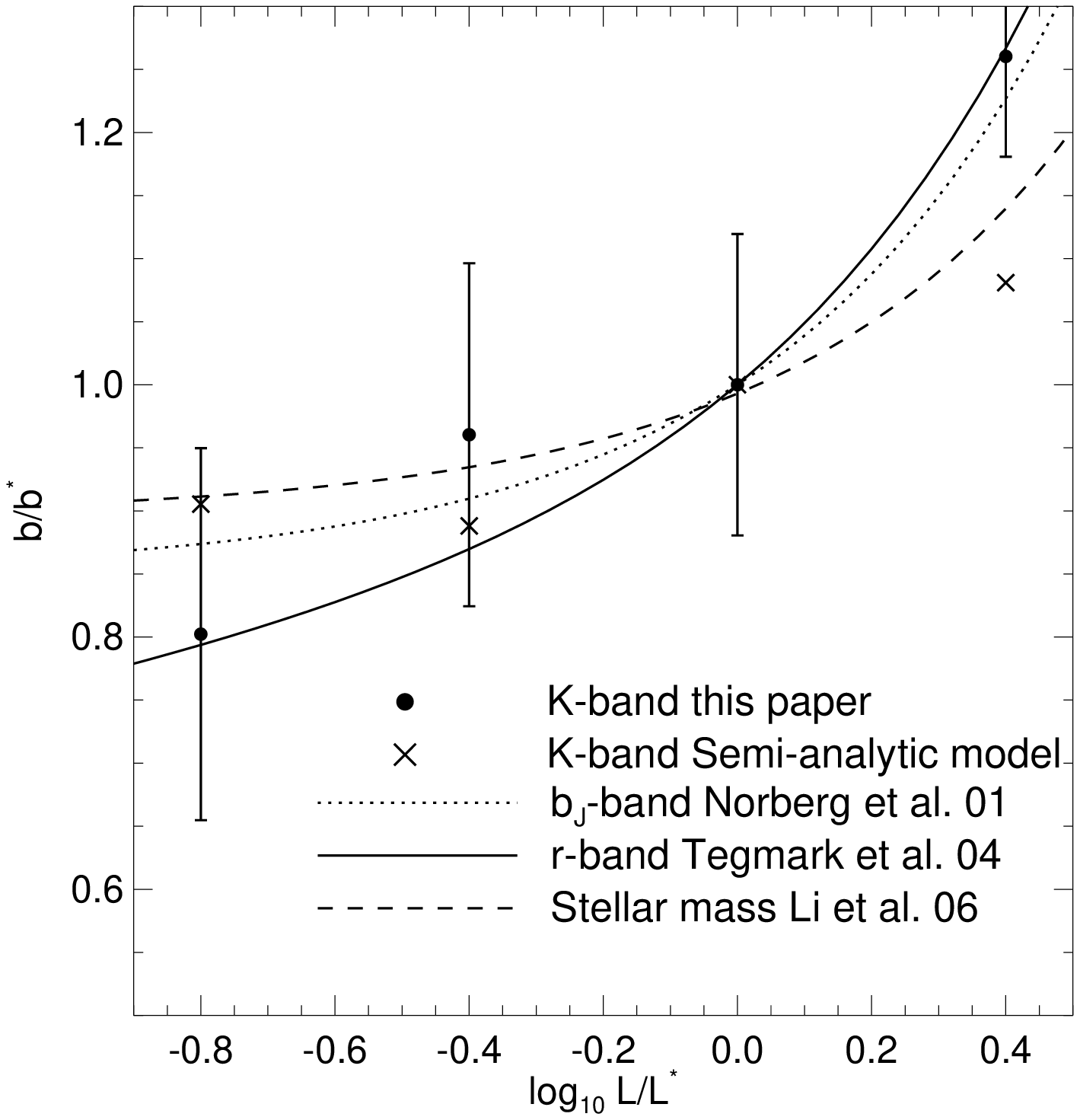}}
\caption{Top: projected correlation functions $w_p(\sigma)$ for the
volume-limited sub-samples. Bottom: luminosity dependence of the
comoving correlation lengths and the linear bias parameters of different
galaxy samples and the mock galaxy sample.
The relative bias factors $b/b^*$ is defined by the amplitude
of $w_p(\sigma)$ measured at $\sigma = 2.7\hmpc$
divided by the same quantity of the $-24 < M_K < -23$ sub-sample
($L \approx L^*$), the dotted curve is at the $b_J$-band of the 2dFGRS survey
$b/b^* = 0.85 + 0.15L/L^*$ \citep{NorbergEtal2001}, the solid curve is
from the SDSS $r$-band power spectrum $b/b^* = 0.85 +
0.15L/L^* - 0.04(M- M^*$) \citep{TegmarkEtal2004b}, and the dashed curve is
a fit to the $w_p(\sigma)$ of stellar mass sub-samples
$b/b^* = 0.896 + 0.097M/M^*$ \citep{LiEtal2006} in which $M$ and
$M^*$ refer to stellar mass.}
\label{fig:Wp_lum}
\end{figure}

Figure~\ref{fig:Wp_lum} shows the projected correlation functions
$w_p(\sigma)$ of volume-limited samples, and Table~\ref{tab:vl}
lists the parameters $r_0$ and $\gamma$ for power law $\xi(r)$
models by fitting $w_p(r_p)$ in the range $0.1\hmpc < \sigma <
13\hmpc$. The slopes of these sub-samples are roughly constant with
errorbars, indicating weak dependence on luminosity; but the
correlation length $r_0$ does increase with luminosity, thus being
consistent with earlier studies \citep{NorbergEtal2001,
ZehaviEtal2005, LiEtal2006}, and the $r_0$ -- $L$ relation follows
the same law as  the relations in other bands and simulations
(Figure~\ref{fig:Wp_lum}).

We also explore the relative bias factor $b/b^*$ computed using the
ratio of the $w_p(2.7\hmpc)$ of our four sub-samples to the
$w_p(2.7\hmpc)$ of the $-24 < M_K < -23$ sub-sample ($L \approx
L^*$). This fiducial separation of $2.7\hmpc$ is chosen because it
is well out of the extremely non-linear regime, but still small
enough so that the correlation functions are very accurately
measured for all sub-samples. In the bottom right panel of
Figure~\ref{fig:Wp_lum}, we compared our measured $b/b^*$ to the
models of \citet{NorbergEtal2001}, \citet{TegmarkEtal2004b} and
\citet{LiEtal2006} together with data from simulations. Within
the estimated uncertainties, the models of \citet{NorbergEtal2001} and
\citet{TegmarkEtal2004b} are basically consistent with our
estimations, but the model of \citet{LiEtal2006} and the
semi-analytical model fail to produce enough clustering power in the
highest luminosity bin.

\begin{figure}
\begin{center}
\resizebox{0.7\hsize}{0.6\hsize}{\includegraphics{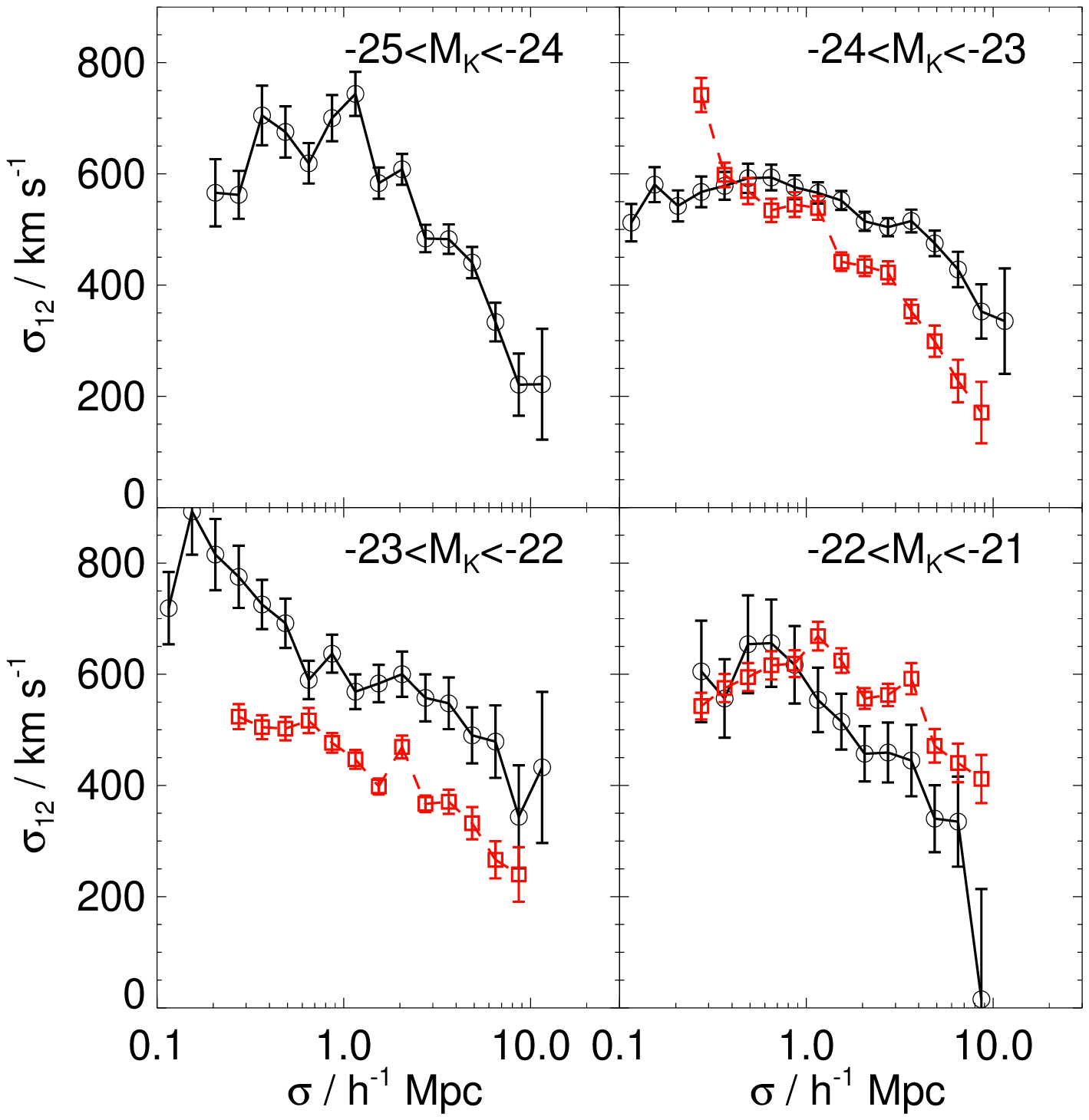}}
\end{center}
\caption{$\sigma_{12}$ of volume limited sub-samples (circles) against
semi-analytical modelling (squares).}
\label{fig:vl_pvd}
\end{figure}

The luminosity dependence of $\sigma_{12}$ is rather much
complicated as already realized by \citet{JingBorner2004b} and
\citet{LiEtal2006b}. It is pointed out that, at small separation,
$\sigma_{12}$ has a trough at $L^*$, becomes relative flat at lower
luminosity end and increasing rapidly at $L>L^*$. This has not been
reproduced by any current halo models. Figure~\ref{fig:vl_pvd} shows
that the $\sigma_{12}$ curve of $L^*$ sub-sample at $\sigma<1\hmpc$
is fairly flat and has smaller amplitude than other luminosity
sub-samples, implying a relative smaller fraction of satellites
within the luminosity bin or the stellar mass range. Interestingly
that $\sigma_{12}$ curves of the brightest sub-sample and the
faintest sub-sample have very similar shape although amplitudes
differ slightly, it seems that the two sub-samples contain similar
fraction of satellites although their occupied halos have very
different masses, since the brightest sub-sample has a much bigger
bias than the faintest sub-sample.

The semi-analytical model works well
for galaxies with low K-band luminosity, but
displays significant discrepancies at high luminosity end
despite the rough agreement in spatial clustering (Table~\ref{tab:vl}). The
case in K-band is different to the discovery of \citet{LiEtal2007} at $r$-band where there are less
differences between the semi-analytical model and the data at lower luminosity end.
However the inconsistency of the semi-analytical model may be cuased by the limited size of
the simulation,  henceforth deficiency in large mass halos to provide sufficient number of
galaxies of high stellar masses.

\section{summary and discussion}
For furnishing the clustering evolution of NIR galaxies, a local
galaxy catalogue limited by K-band magnitudes of $10<K<13.5$ is generated
by cross-matching the 2MASS data with the SDSS survey,
then we carefully measured the 2PCFs and the pairwise velocity dispersion of
the galaxy sample, as well as their color and luminosity dependence respectively.

In the redshift space, the 2PCF $\xi(s)$ of the flux-limited sample
complies with shallower power law at scales of $s<\sim 3\hmpc$ than
that at larger scales, but in real space
the 2PCF $\xi(r)$ derived from the projected 2PCF $w(\sigma)$
is perfectly approximated by a single power law with $r_0=6.44\pm0.23\hmpc$
and $\gamma=1.81\pm0.02$ for $r<1\sim 10\hmpc$ (Figure~\ref{fig:Wp}). The
estimated correlation length supports the conclusion of \citet{WaddingtonEtal2007}
that the clustering of NIR galaxies evolves very slow upto $z\sim 0.5$.
The pairwise velocity dispersion at small scales is directly
related to the spatial distribution in the halo \citep{SlosarEtal2006},
our $\sigma_{12}$ of local NIR galaxies
has shape analogous to those found in simulations, showing a bump
at scales around $1\hmpc$, but is larger than in optical bands, being
$685\pm 17\kms$ if assuming scale invariant $\sigma_{12}$ (Table~\ref{table:comp}).
The phenomenon could be rooted in different luminosity functions
at these bands, it is likely that the distribution of $K$-band luminosity or
stellar mass of satellite galaxies is more concentrated in the range defining our
sample than that of central galaxies. Regrettably we do not have
a large NIR galaxy sample at high redshift with redshift, otherwise
comparison of $\sigma_{12}$ would enable us peering into
the evolution of the positioning of NIR galaxies in their host halos,
or the velocity biasing of galaxies relative to the dark matter.

The optical color dependence of the clustering of NIR galaxies
is similar to optical bands selected galaxies,  blue galaxies
of $g-r <0.7$ are much less clustered than red galaxies, and
displaying a very flat and low $\sigma_{12}$ in nonlinear regime
as consequences of having smaller fraction of satellite galaxies.
And the luminosity dependence of 2PCF and $\sigma_{12}$ of our NIR
galaxies is not different with that of optical galaxies in spite of
the very different luminosity functions. As the $K$-band luminosity
is tightly correlated with the stellar mass, it is
somehow surprising to discover that the empirical formula of relative biasing
by stellar mass in \citet{LiEtal2006} under-predict the bias
of the brightest volume-limited sample. Also, the transition
of the shape and amplitude of $\sigma_{12}$ at $L^*$ might infer that
the luminosity distribution of satellite galaxies in low mass halos
are very different to that in high mass halos.

Examination of our galaxy sample against mock galaxy sample
reveals that the $K$-band luminosity dependence of the 2PCF
can be approximately reproduced by semi-analytical modelling within
measurement errors, but the $\sigma_{12}$ of the mock
deviates from observation significantly in aspects of amplitude, shape and
luminosity dependence, especially for bright galaxies.
The peculiar velocity of galaxy
is a conundrum in galaxy formation models both in theories and simulations,
accurate modelling requires exquisite fabrication of galaxies in halos and
corresponding evolution paths, it seems so far
there is still a long way to go.

\begin{acknowledgements}
This work is funded by the NSFC under grants of
Nos.10643002, 10633040, 10873035, 10725314 and the Ministry of Science \& Technology
of China through 973 grant of No. 2007CB815402.
We thank Xi Kang for providing the mock galaxy catalogue.
The N-body simulation was performed at Shanghai Supercomputer
Center by Weipeng Lin under the financial support of Chinese
863 project (No. 2006AA01A125).

This publication makes use of data products from the {\it Two Micron
All Sky Survey}, which is a joint project of the University of
Massachusetts and the Infrared Processing and Analysis
Center/California Institute of Technology, funded by the National
Aeronautics and Space Administration and the National Science
Foundation.

This publication also makes use of the {\it Sloan
Digital Sky Survey} (SDSS). Funding for the creation and
distribution of the SDSS Archive has been provided by the Alfred P.\
Sloan Foundation, the Participating Institutions, the National
Aeronautics and Space Administration, the National Science
Foundation, the US Department of Energy, the Japanese
Monbukagakusho, and the Max Planck Society.  The SDSS Web site is
\texttt {http://www.sdss.org/}.  The SDSS Participating Institutions
are the University of Chicago, Fermilab, the Institute for Advanced
Study, the Japan Participation Group, the Johns Hopkins University,
the Max Planck Institut f\"ur Astronomie, the Max Planck Institut
f\"ur Astrophysik, New Mexico State University, Princeton
University, the United States Naval Observatory, and the University
of Washington. This publication also made use of NASA's Astrophysics
Data System Bibliographic Services.
\end{acknowledgements}


\begin{thebibliography}{46}
\expandafter\ifx\csname natexlab\endcsname\relax\def\natexlab#1{#1}\fi

\bibitem[{{Adelman-McCarthy} {et~al.}(2008)}]{Adelman-McCarthyEtal2008}
{Adelman-McCarthy}, J.~K., {et~al.} 2008, \apjs, 175, 297

\bibitem[{{Baugh}(2006)}]{Baugh2006}
{Baugh}, C.~M. 2006, Reports on Progress in Physics, 69, 3101

\bibitem[{{Baugh} {et~al.}(1996){Baugh}, {Gardner}, {Frenk}, \&
  {Sharples}}]{BaughEtal1996}
{Baugh}, C.~M., {Gardner}, J.~P., {Frenk}, C.~S., \& {Sharples}, R.~M. 1996,
  \mnras, 283, 15

\bibitem[{{Bell} {et~al.}(2003){Bell}, {McIntosh}, {Katz}, \&
  {Weinberg}}]{BellEtal2003}
{Bell}, E.~F., {McIntosh}, D.~H., {Katz}, N., \& {Weinberg}, M.~D. 2003, \apjs,
  149, 289

\bibitem[{{Bernardeau} {et~al.}(2002){Bernardeau}, {Colombi}, {Gazta{\~n}aga},
  \& {Scoccimarro}}]{BernardeauEtal2002}
{Bernardeau}, F., {Colombi}, S., {Gazta{\~n}aga}, E., \& {Scoccimarro}, R.
  2002, \physrep, 367, 1

\bibitem[{{Blanton} {et~al.}(2005)}]{BlantonEtal2005}
{Blanton}, M.~R., {et~al.} 2005, \aj, 129, 2562

\bibitem[{{Carlberg} {et~al.}(1997){Carlberg}, {Cowie}, {Songaila}, \&
  {Hu}}]{CarlbergEtal1997}
{Carlberg}, R.~G., {Cowie}, L.~L., {Songaila}, A., \& {Hu}, E.~M. 1997, \apj,
  484, 538

\bibitem[{{Cooray} \& {Sheth}(2002)}]{CooraySheth2002}
{Cooray}, A., \& {Sheth}, R. 2002, \physrep, 372, 1

\bibitem[{{Cowie} {et~al.}(1994){Cowie}, {Gardner}, {Hu}, {Songaila}, {Hodapp},
  \& {Wainscoat}}]{CowieEtal1994}
{Cowie}, L.~L., {Gardner}, J.~P., {Hu}, E.~M., {Songaila}, A., {Hodapp}, K.-W.,
  \& {Wainscoat}, R.~J. 1994, \apj, 434, 114

\bibitem[{{Davis} \& {Peebles}(1983)}]{DavisPeebles1983}
{Davis}, M., \& {Peebles}, P.~J.~E. 1983, \apj, 267, 465

\bibitem[{{Efstathiou}(1988)}]{Efstathiou1988}
{Efstathiou}, G. 1988, in Lecture Notes in Physics, Berlin Springer Verlag,
  Vol. 297, Comets to Cosmology, ed. A.~{Lawrence}, 312

\bibitem[{{Hamilton}(1992)}]{Hamilton1992}
{Hamilton}, A.~J.~S. 1992, \apj, 385, 5

\bibitem[{{Hamilton}(1993)}]{Hamilton1993}
{Hamilton}, A.~J.~S. 1993, \apj, 417, 19

\bibitem[{{Hamilton}(1998)}]{Hamilton1998}
{Hamilton}, A.~J.~S. 1998, in ASSL Vol. 231: The Evolving Universe, 185

\bibitem[{{Hawkins} {et~al.}(2003)}]{HawkinsEtal2003}
{Hawkins}, E., {et~al.} 2003, \mnras, 346, 78

\bibitem[{{Jarrett} {et~al.}(2000){Jarrett}, {Chester}, {Cutri}, {Schneider},
  {Skrutskie}, \& {Huchra}}]{JarrettEtal2000}
{Jarrett}, T.~H., {Chester}, T., {Cutri}, R., {Schneider}, S., {Skrutskie}, M.,
  \& {Huchra}, J.~P. 2000, \aj, 119, 2498

\bibitem[{{Jing} \& {B{\"o}rner}(2004)}]{JingBorner2004b}
{Jing}, Y.~P., \& {B{\"o}rner}, G. 2004, \apj, 617, 782

\bibitem[{{Kaiser}(1987)}]{Kaiser1987}
{Kaiser}, N. 1987, \mnras, 227, 1

\bibitem[{{Kang} {et~al.}(2005){Kang}, {Jing}, {Mo}, \&
  {B{\"o}rner}}]{KangEtal2005}
{Kang}, X., {Jing}, Y.~P., {Mo}, H.~J., \& {B{\"o}rner}, G. 2005, \apj, 631, 21

\bibitem[{{Landy} \& {Szalay}(1993)}]{LandySzalay1993}
{Landy}, S.~D., \& {Szalay}, A.~S. 1993, \apj, 412, 64

\bibitem[{{Li} {et~al.}(2007){Li}, {Jing}, {Kauffmann}, {B{\"o}rner}, {Kang},
  \& {Wang}}]{LiEtal2007}
{Li}, C., {Jing}, Y.~P., {Kauffmann}, G., {B{\"o}rner}, G., {Kang}, X., \&
  {Wang}, L. 2007, \mnras, 376, 984

\bibitem[{{Li} {et~al.}(2006{\natexlab{a}}){Li}, {Jing}, {Kauffmann},
  {B{\"o}rner}, {White}, \& {Cheng}}]{LiEtal2006b}
{Li}, C., {Jing}, Y.~P., {Kauffmann}, G., {B{\"o}rner}, G., {White}, S.~D.~M.,
  \& {Cheng}, F.~Z. 2006{\natexlab{a}}, \mnras, 368, 37

\bibitem[{{Li} {et~al.}(2006{\natexlab{b}}){Li}, {Kauffmann}, {Jing}, {White},
  {B{\"o}rner}, \& {Cheng}}]{LiEtal2006}
{Li}, C., {Kauffmann}, G., {Jing}, Y.~P., {White}, S.~D.~M., {B{\"o}rner}, G.,
  \& {Cheng}, F.~Z. 2006{\natexlab{b}}, \mnras, 368, 21

\bibitem[{{Lin} {et~al.}(2006){Lin}, {Jing}, {Mao}, {Gao}, \&
  {McCarthy}}]{LinEtal2006}
{Lin}, W.~P., {Jing}, Y.~P., {Mao}, S., {Gao}, L., \& {McCarthy}, I.~G. 2006,
  \apj, 651, 636

\bibitem[{{Lupton}(1993)}]{Lupton1993}
{Lupton}, R. 1993, {Statistics in theory and practice} (Princeton, N.J.,
  Princeton University Press)

\bibitem[{{McCracken} {et~al.}(2000){McCracken}, {Shanks}, {Metcalfe}, {Fong},
  \& {Campos}}]{McCrackenEtal2000}
{McCracken}, H.~J., {Shanks}, T., {Metcalfe}, N., {Fong}, R., \& {Campos}, A.
  2000, \mnras, 318, 913

\bibitem[{{Mobasher} {et~al.}(1986){Mobasher}, {Ellis}, \&
  {Sharples}}]{MobasherEtal1986}
{Mobasher}, B., {Ellis}, R.~S., \& {Sharples}, R.~M. 1986, \mnras, 223, 11

\bibitem[{{Norberg} {et~al.}(2001){Norberg}, {Baugh}, {Hawkins}, {Maddox},
  {et~al.}}]{NorbergEtal2001}
{Norberg}, P., {Baugh}, C.~M., {Hawkins}, E., {Maddox}, S., {et~al.} 2001,
  \mnras, 328, 64

\bibitem[{{Peebles}(1980)}]{Peebles1980}
{Peebles}, P.~J.~E. 1980, {The large-scale structure of the universe}
  (Princeton, N.J., Princeton University Press)

\bibitem[{{Peebles}(1993)}]{Peebles1993}
{Peebles}, P.~J.~E. 1993, {Principles of physical cosmology}
  (Princeton, N.J., Princeton University Press)

\bibitem[{{Phillipps} {et~al.}(1978){Phillipps}, {Fong}, {Fall}, \&
  {MacGillivray}}]{PhillippsEtal1978}
{Phillipps}, S., {Fong}, R., {Fall}, R.~S.~E.~S.~M., \& {MacGillivray}, H.~T.
  1978, \mnras, 182, 673

\bibitem[{{Roche} {et~al.}(2003){Roche}, {Dunlop}, \&
  {Almaini}}]{RocheEtal2003}
{Roche}, N.~D., {Dunlop}, J., \& {Almaini}, O. 2003, \mnras, 346, 803

\bibitem[{{Sandage} {et~al.}(1979){Sandage}, {Tammann}, \& {Yahil}}]{STY1979}
{Sandage}, A., {Tammann}, G.~A., \& {Yahil}, A. 1979, \apj, 232, 352

\bibitem[{{Saunders} {et~al.}(1992){Saunders}, {Rowan-Robinson}, \&
  {Lawrence}}]{SaundersEtal1992}
{Saunders}, W., {Rowan-Robinson}, M., \& {Lawrence}, A. 1992, \mnras, 258, 134

\bibitem[{{Schechter}(1976)}]{Schechter1976}
{Schechter}, P. 1976, \apj, 203, 297

\bibitem[{{Scoccimarro}(2004)}]{Scoccimarro2004}
{Scoccimarro}, R. 2004, \prd, 70, 083007

\bibitem[{{Slosar} {et~al.}(2006){Slosar}, {Seljak}, \&
  {Tasitsiomi}}]{SlosarEtal2006}
{Slosar}, A., {Seljak}, U., \& {Tasitsiomi}, A. 2006, \mnras, 366, 1455

\bibitem[{{Springel}(2005)}]{Springel2005}
{Springel}, V. 2005, \mnras, 364, 1105

\bibitem[{{Tegmark} {et~al.}(2004)}]{TegmarkEtal2004b}
{Tegmark}, M., {et~al.} 2004, \apj, 606, 702

\bibitem[{{Waddington} {et~al.}(2007){Waddington}, {Oliver}, {Babbedge},
  {Fang}, {Farrah}, {Franceschini}, {Gonzalez-Solares}, {Lonsdale},
  {Rodighiero}, {Rowan-Robinson}, {Shupe}, {Surace}, {Vaccari}, \&
  {Xu}}]{WaddingtonEtal2007}
{Waddington}, I., {et~al.} 2007, \mnras, 381, 1437

\bibitem[{{Weinberg} {et~al.}(2004){Weinberg}, {Dav{\'e}}, {Katz}, \&
  {Hernquist}}]{WeinbergEtal2004}
{Weinberg}, D.~H., {Dav{\'e}}, R., {Katz}, N., \& {Hernquist}, L. 2004, \apj,
  601, 1

\bibitem[{{York} {et~al.}(2000)}]{YorkEtal2000}
{York}, D.~G., {et~al.} 2000, \aj, 120, 1579

\bibitem[{{Yoshikawa} {et~al.}(2003){Yoshikawa}, {Jing}, \&
  {B{\"o}rner}}]{YoshikawaEtal2003}
{Yoshikawa}, K., {Jing}, Y.~P., \& {B{\"o}rner}, G. 2003, \apj, 590, 654

\bibitem[{{Zehavi} {et~al.}(2002){Zehavi}, {Blanton}, {Frieman}, {Weinberg},
  {Mo}, {Strauss}, {Anderson}, {Annis}, {et~al.}}]{ZehaviBlantonEtal2002}
{Zehavi}, I., {et~al.} 2002, \apj,
  571, 172

\bibitem[{{Zehavi} {et~al.}(2005)}]{ZehaviEtal2005}
{Zehavi}, I., {et~al.} 2005, \apj, 630, 1

\bibitem[{{Zhao} {et~al.}(2002){Zhao}, {Jing}, \& {B{\"o}rner}}]{ZhaoEtal2002}
{Zhao}, D., {Jing}, Y.~P., \& {B{\"o}rner}, G. 2002, \apj, 581, 876

\end{thebibliography}

\end{document}